\documentclass{article}
\usepackage{arxiv}
\usepackage{subfigure}
\usepackage[utf8]{inputenc} 
\usepackage[T1]{fontenc}    
\usepackage{hyperref}       
\usepackage{url}            
\usepackage{booktabs}       
\usepackage{amsfonts}       
\usepackage{nicefrac}       
\usepackage{microtype}      
\usepackage{lipsum}		
\usepackage{graphicx}
\usepackage{natbib}
\usepackage{doi}
\usepackage{algorithm}
\usepackage{algpseudocode}
\usepackage{amsmath}
\usepackage{xcolor}
\usepackage{verbatim}
\usepackage{array}
\usepackage{setspace}
\usepackage{wasysym}

\title{Introduction to Eye Tracking: A Hands-On Tutorial for Students and Practitioners}

\author{ \href{https://orcid.org/0000-0003-3146-4484}{\includegraphics[scale=0.06]{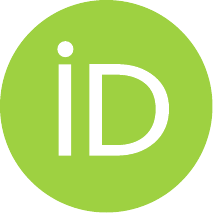}\hspace{1mm}Enkelejda Kasneci} \\
	Technical University of Munich\\
	\texttt{enkelejda.kasneci@tum.de} \\
        \And
	\href{https://orcid.org/0000-0003-3934-433X}{\includegraphics[scale=0.06]{orcid.pdf}\hspace{1mm}Hong Gao} \\
	Technical University of Munich \\
	\texttt{hong.gao@tum.de} \\
        \And
	\href{https://orcid.org/0000-0002-3390-6154}{\includegraphics[scale=0.06]{orcid.pdf}\hspace{1mm}Suleyman Ozdel} \\
	Technical University of Munich \\
	\texttt{ozdelsuleyman@tum.de} \\
        \And
	{Virmarie Maquiling} \\
	Technical University of Munich \\
	\texttt{virmarie.maquiling@tum.de} \\
        \And
	\href{https://orcid.org/0009-0007-5111-3776}{\includegraphics[scale=0.06]{orcid.pdf}\hspace{1mm}Enkeleda Thaqi} \\
	Technical University of Munich \\
	\texttt{enkeleda.thaqi@tum.de} \\
        \And
	{Carrie Lau} \\
	Technical University of Munich \\
	\texttt{carrie.lau@tum.de} \\
        \And
	\href{https://orcid.org/0000-0002-6031-3741}{\includegraphics[scale=0.06]{orcid.pdf}\hspace{1mm}Yao Rong} \\
	Technical University of Munich \\
	\texttt{yao.rong@tum.de} \\
        \And
    \href{https://orcid.org/0000-0002-3123-7268}{\includegraphics[scale=0.06]{orcid.pdf}\hspace{1mm}Gjergji Kasneci} \\
	Technical University of Munich \\
	\texttt{gjergji.kasneci@tum.de} \\
        \And
	\href{https://orcid.org/0000-0002-4594-4318}{\includegraphics[scale=0.06]{orcid.pdf}\hspace{1mm}Efe Bozkir} \\
	Technical University of Munich \\
	\texttt{efe.bozkir@tum.de} \\
}

\renewcommand{\shorttitle}{A Hands-on Tutorial for Eye Tracking}

\hypersetup{
pdftitle={Introduction to Eye Tracking: A Hands-On Tutorial for Students and Practitioners},
pdfsubject={Eye-tracking tutorial},
pdfauthor={Enkelejda Kasneci, Hong Gao, Suleyman Ozdel, Virmarie Maquiling,  Enkeleda Thaqi, Carrie Lau, Yao Rong, Gjergji Kasneci, Efe Bozkir},
pdfkeywords={Eye tracking, Eye movements, Scanpaths, Data processing, User studies},
}

\begin{document}
\maketitle

\begin{abstract}
Eye-tracking technology is widely used in various application areas such as psychology, neuroscience, marketing, and human-computer interaction, as it is a valuable tool for understanding how people process information and interact with their environment. This tutorial provides a comprehensive introduction to eye tracking, from the basics of eye anatomy and physiology to the principles and applications of different eye-tracking systems.
The guide is designed to provide a hands-on learning experience for everyone interested in working with eye-tracking technology. Therefore, we include practical case studies to teach students and professionals how to effectively set up and operate an eye-tracking system. The tutorial covers a variety of eye-tracking systems, calibration techniques, data collection, and analysis methods, including fixations, saccades, pupil diameter, and visual scan path analysis. 
In addition, we emphasize the importance of considering ethical aspects when conducting eye-tracking research and experiments, especially informed consent and participant privacy. We aim to give the reader a solid understanding of basic eye-tracking principles and the practical skills needed to conduct their experiments. 
Python-based code snippets and illustrative examples are included in the tutorials and can be downloaded at: \url{https://gitlab.lrz.de/hctl/Eye-Tracking-Tutorial}.
\end{abstract}

\keywords{Eye tracking \and Eye movements \and Scanpaths \and Data processing \and User studies}

\newpage
  \tableofcontents
\newpage

\section{Introduction to Eye Tracking}
A basic comprehension of eye anatomy and physiology provides the foundation for the understanding and effective use of eye-tracking technology. Hence, in the following, we introduce eye-tracking technology by focusing first on the anatomy and physiology of the eye, and provide then an overview of basic eye-tracking techniques and current types of eye trackers.

\subsection{Basics of Eye Anatomy and Physiology}
\begin{figure}[h]
\centering
\includegraphics[width=.5\linewidth]{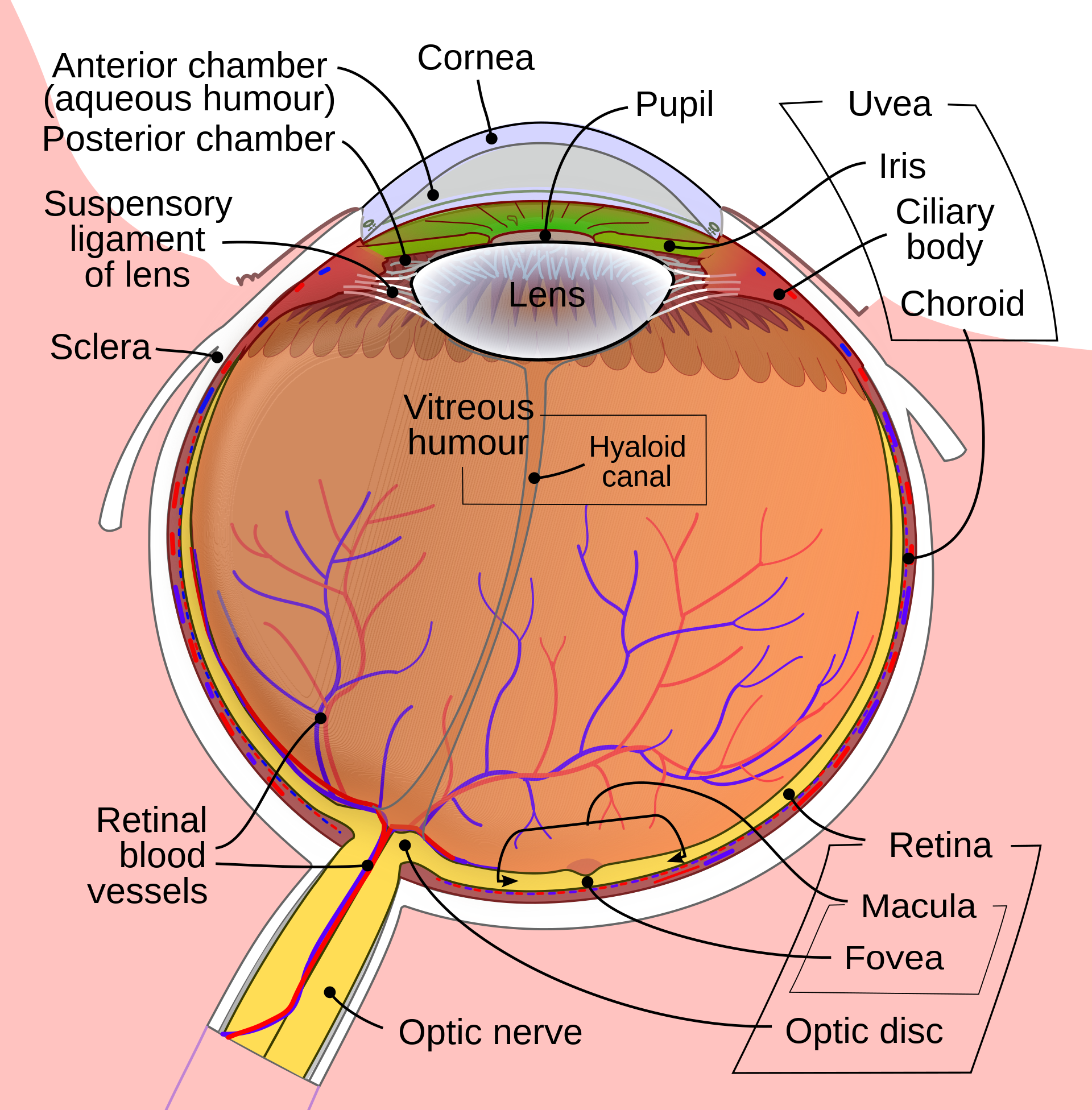}
\caption{Human Eye~\citep{humanEyeImage}.}
\label{fig:HumanEye}
\end{figure}

The human eye, its anatomy and physiology are the basis for our visual perception. Its main task is to capture incoming light and convert it into electrical signals. These signals are then transmitted to the brain, where they are interpreted and result in the information of a visual image.  Figure~\ref{fig:HumanEye} illustrates the different key components of the human eye, including the cornea, iris, pupil, lens, retina, and optic nerve. In the following, we offer a concise overview of the basic anatomy and physiology of the eye, with a specific focus on features relevant to eye tracking \citep{atchison2023optics, holmqvist2011eye, kolb1995eye}.

The cornea, a transparent structure, is the outermost layer of the eye and covers both the iris and the pupil. One of its most important functions is to refract incoming light and direct it onto the retina, playing a crucial role in vision. In addition, the cornea acts as a protective barrier against external agents such as dust or debris.

The iris, the colored part of the eye, regulates the amount of light entering the eye and controls the size of the pupil. The pupil itself can dilate or constrict in response to light intensity. Positioned behind the iris, the lens fine-tunes the focus of incoming light. The curvature of the lens can be altered by the ciliary muscles, which adjust its shape to allow for accommodation, the process of changing focus between distant and near objects \citep{singh2012humaneye}.

The retina, positioned at the rear of the eye, is a light-sensitive tissue composed of multiple layers of neurons and photoreceptors. These photoreceptors, consisting of rods and cones, convert incoming light into neural signals. Rods are responsible for detecting light, while cones are responsible for perceiving color. Within the retina, the macula - a small central area - contains a dense concentration of cones. This area is responsible for high visual acuity, color vision, and detail perception \citep{koretz1988humaneye}.

The optic nerve, originating from the back of the retina, transmits visual information to the brain for processing and interpretation. The optic nerve comprises millions of axons, bundled together to form a compact structure that exits the eye and carries information to several brain regions, including the visual cortex, responsible for creating the conscious perception of visual information \citep{atchison2023optics}.

The complex movement of the eye is controlled by a series of six extraocular muscles that control the position of the eye, as well as the levator palpebrae superioris muscle, which controls the elevation of the eyelid. These muscles are innervated by several cranial nerves and coordinate to allow precise eye movements, such as smooth pursuit or saccades, which are crucial to exploring and analyzing the visual environment \citep{singh2012humaneye}.

The movement of our eyes is regulated by several brain regions, including the brainstem and the cerebellum, which work together to ensure coordinated eye movements and stable images on the retina. These regions integrate information from the vestibular system, responsible for detecting head movements and maintaining balance, and the visual system, allowing for a precise and rapid response to changes in the visual environment.

As described above, the anatomy and physiology of the eye is a complex network of structures and processes responsible for vision, eye movements and visual perception \citep{atchison2023optics}. Understanding the basic principles of eye anatomy and physiology is essential for understanding the potential and applications of eye-tracking technologies in research and clinical settings. 

\subsection{Basic Concepts in Eye-tracking Metrics}
\subsubsection{Sampling rate}
The sampling frequency of an eye-tracking system refers to how many times per second the eye tracker registers the position of the eyes, which is measured in Hertz (Hz)~\citep{andersson2010sampling}. The higher the sampling frequency, the more precise and accurate the eye tracker's ability to estimate the true movements of the eyes. A higher sampling rate is particularly important in dynamic environments or tasks that involve quick eye movements, ensuring that no detailed data points are missed. However, a higher sampling frequency also comes at a cost due to the need for advanced cameras, processing capabilities, and more required data storage, etc. When choosing an eye-tracking system for research, it is important to find a balance between the desired sampling frequency the available budget, and the specific research objectives.

\subsubsection{Accuracy and precision}
\textbf{Accuracy} in eye-tracking is the deviation between the actual gaze position and the gazed position recorded by the eye tracker, i.e., how closely the recorded eye-tracking data matches the actual eye-tracking data. Factors influencing accuracy include calibration quality, camera resolution, and the stability of the eye-tracking system. High accuracy ensures reliable and trustworthy data, which is crucial for tasks requiring precise eye movement analysis, for example clinical assessments.

\begin{figure}[h]
    \centering
\includegraphics[width=.7\linewidth]{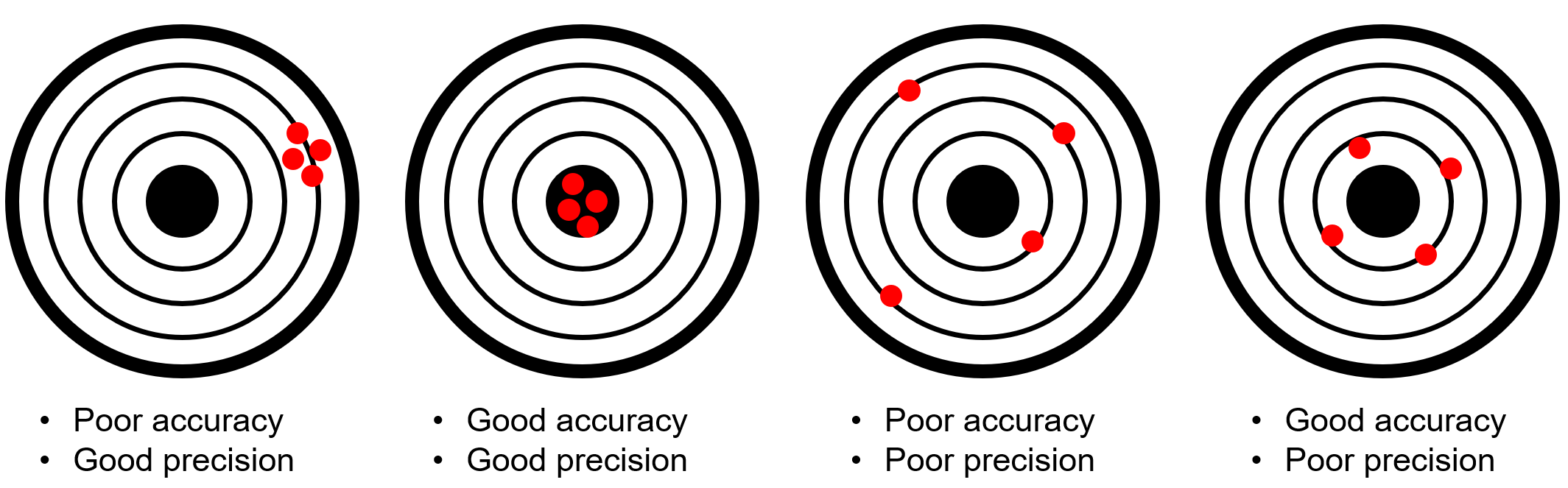}
    \caption{Precision and accuracy in eye-tracking systems.}
    \label{fig:pre-acc}
\end{figure}

\textbf{Precision} refers to the measure of variation in the recorded data.  It measures the variability in tracking the same eye movement multiple times under the same conditions, which is calculated using the Root Mean Square (RMS) of the sampled points. A high-precision eye-tracking system will produce similar results when tracking the same eye movement repeatedly, indicating low variability or error in the measurements. Precision is essential for obtaining reliable and consistent data, especially in research contexts where subtle differences in eye movements need to be detected and analyzed accurately. See an illustration in Figure~\ref{fig:pre-acc}

\subsection{Eye-tracking Techniques and Current Types of Eye-tracking Systems}

This section offers an overview of the different types of eye-tracking systems, including remote and head-mounted eye trackers, along with their respective strengths and limitations.

Eye tracking is a technique used to objectively measure and record the direction of an individual's gaze and eye movements~\citep{wade2005eye}. These are accomplished by measuring the relative position of the eye in relation to the head or the orientation of the gaze itself~\citep{duchowksi2007eye}. Subsequently, collected data can be analyzed and evaluated in various ways, such as identifying eye movement patterns or identifying frequently viewed areas of interest~\citep{holmqvist2011eye}. 

\begin{table}[t]
\centering
\caption{Techniques for eye movement recording}
\label{table:tech_emrecord}
\begin{tabular}{p{0.15\linewidth}p{0.26\linewidth}p{0.25\linewidth}p{0.25\linewidth}}
        \toprule
        \textbf{Method} & \textbf{Description} & \textbf{Advantages} & \textbf{Disadvantages} \\
        \midrule
        \textbf{Electro-oculography (EOG)} & Uses electrodes placed around the eyes to detect electrical signals generated by eye muscles. & Non-invasive, can detect eye movements even when eyes are closed. & Limited spatial resolution, sensitive to artifacts from eye blinks and other muscle movements. \\
        \midrule
        \textbf{Scleral Contact Lenses} & Lenses placed directly on the cornea to measure changes in corneal curvature. & High spatial resolution, minimizes artifacts from head movements. & Invasive, may cause discomfort or irritation to the eyes. \\
        \midrule
        \textbf{Video-based Eye Tracking} & Uses a camera directed at the eye to track its movements by analyzing changes in pupil position. & Non-invasive, provides real-time data, suitable for various applications including VR and HCI studies. & May be affected by lighting conditions, requires calibration, limited accuracy for certain eye movements. \\
        \bottomrule
\end{tabular}
\end{table}

There are various methods for recording eye movements (see Table~\ref{table:tech_emrecord}), including electro-oculography, which uses electrodes placed around the eyes to detect electrical signals generated by eye muscles; scleral contact lenses, placed directly on the cornea; or video-based eye tracking, which uses a camera directed at the eye to track its movements~\citep{duchowksi2007eye}. 
Among these methods, video-based eye tracking has gained popularity for most research applications due to its non-invasive nature and minimal influence on eye gaze behavior~\citep{duchowksi2007eye, fuhl2017pupildetection}. 
Most of the remote and head-mounted eye trackers currently available on the market use video-based eye-tracking techniques, which involve a camera and an infrared illumination source. Head-mounted eye trackers, in particular, often include an additional scene camera to capture the surrounding environment~\citep{holmqvist2011eye}. 

The pupil and cornea play a crucial role in video-based eye tracking. Algorithms use measurements of the pupil's center and four corneal reflections relative to the pupil's center to map the gaze points on a screen or scene recording, depending on the specific type of eye tracker employed~\citep{duchowksi2007eye, fuhl2015algo, fuhl2016algo, fuhl2017pupilnet}. Using the resulting gaze points, the respective eye movements can be derived. Corneal reflections serve as additional references for reliable pupil detection and help compensate for minor head movements~\citep{holmqvist2011eye, nitschke2013corneal}. 

\subsubsection{Remote eye tracker} 
\begin{figure}[t]
    \centering
    \includegraphics[width=.7\linewidth]{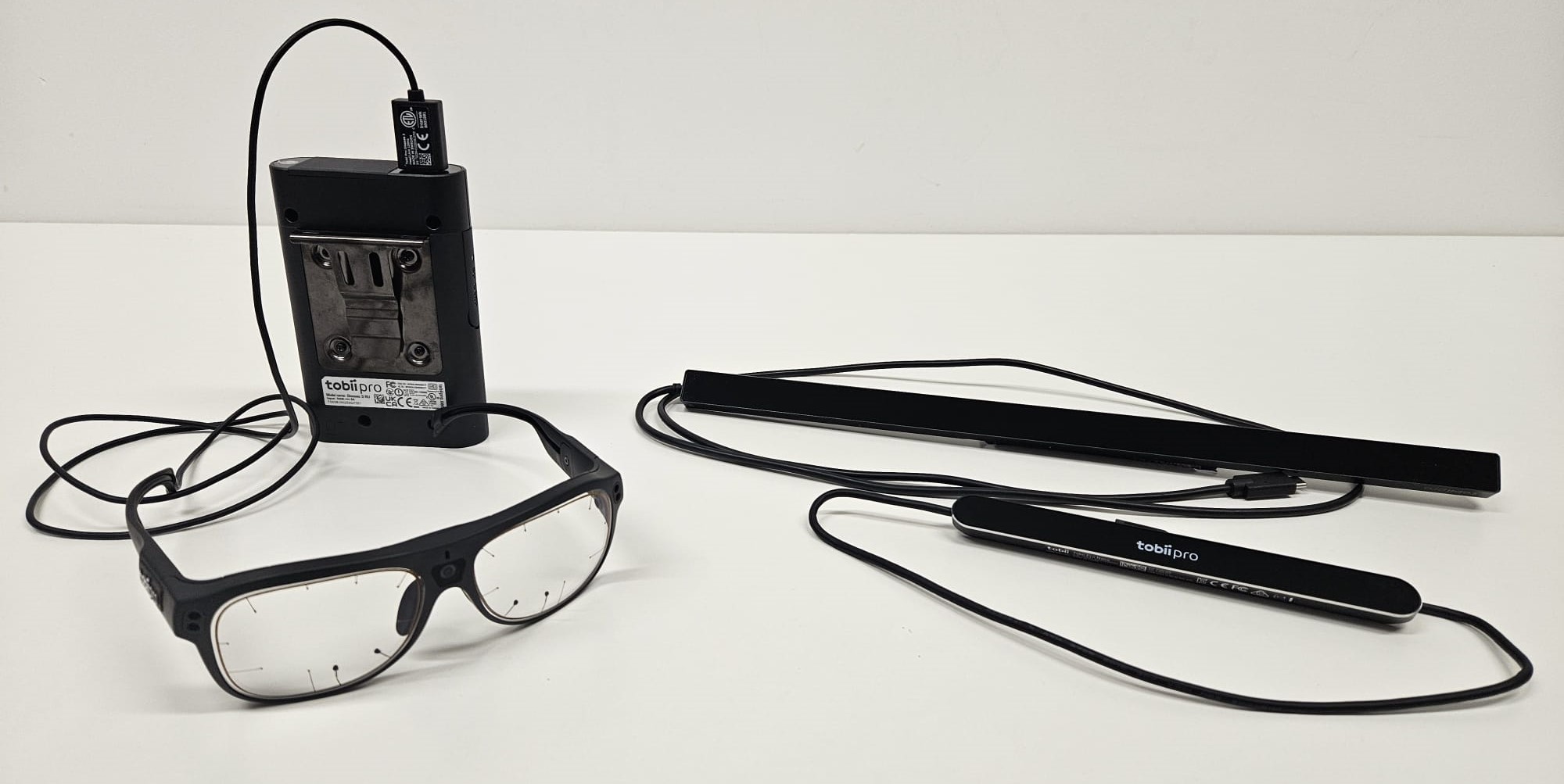} 
    \caption{A wearable eye tracker on the left (Tobii Pro Glasses 3), and remote eye trackers at the top (Tobii Fusion Pro) and bottom (Tobii Pro Nano).}
    \label{fig:eye-tracker}
\end{figure}

Remote eye trackers, for example, manufactured by Tobii, illustrated in Figure \ref{fig:eye-tracker}, are non-invasive systems mounted near a screen to monitor eye movements from a distance. Typically, these systems incorporate a monitor or screen for stimulus presentation and infrared cameras positioned either below or beside the screen to detect the reflection of infrared light from the eyes. Analysis of these reflections, specifically of the cornea and the position of the pupil, enables remote eye trackers to precisely determine where a person is directing their gaze, whether on a screen or within their surrounding environment. 

These systems find common usage in laboratory settings, providing participants with a comfortable seating arrangement while facilitating the recording of their eye movements. A significant advantage of remote eye trackers is their user-friendly nature and adaptability to various experimental setups. Their seamless integration with existing computer setups and software renders them suitable for a broad spectrum of research applications. Moreover, remote eye trackers boast high accuracy and precision in tracking eye movements, facilitating the analysis of subtle changes in gaze behavior. Furthermore, their non-invasive nature minimizes participant discomfort, allowing for extended experimental sessions.

However, remote eye trackers have limitations, particularly in capturing eye movements during dynamic activities or real-world scenarios, as they often require participants to maintain a relatively fixed head position relative to the screen. Additionally, they may encounter interference from occlusions such as glasses, reflections, or obstructions in the participant's field of view, which can disrupt the eye-tracking process and compromise data accuracy. Despite these constraints, remote eye trackers are invaluable for investigating gaze behaviors within controlled laboratory environments. 

\subsubsection{Wearable eye tracker} 
\begin{figure}[b]
    \centering
    \includegraphics[width=.7\linewidth]{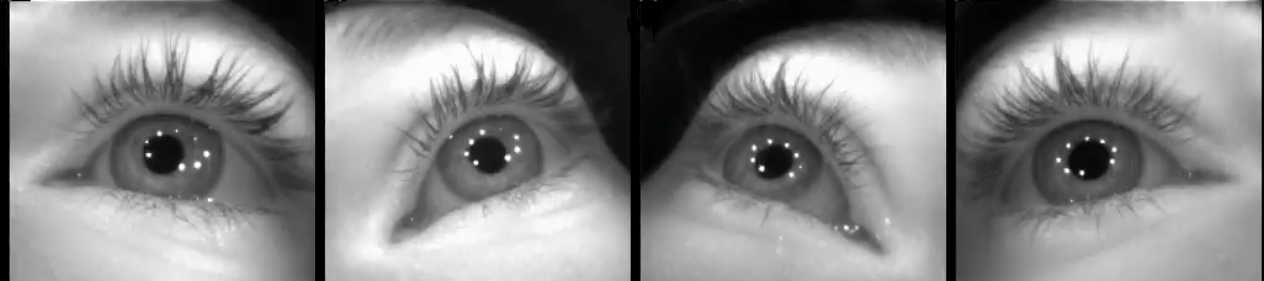} 
    \caption{Eyes' corneal reflections illuminated by infrared light.}
    \label{fig:eye-image}
\end{figure}

Wearable eye trackers from Tobii, depicted in Figure~\ref{fig:eye-tracker}, provide mobility and adaptability by enabling the recording of eye movements in real-world environments. These systems integrate miniature cameras and infrared sensors into lightweight glasses or head-mounted devices worn by the user. Typically, these cameras comprise both eye and scene cameras to capture images of the wearer's eyes and the surrounding environment, while infrared sensors detect reflections from the corneas, enabling precise tracking of eye movements. Figure~\ref{fig:eye-image} depicts eyes being captured simultaneously by two cameras each, with illumination from infrared light to highlight corneal reflections.

These eye-tracking devices are commonly integrated into glasses or head-mounted displays and are gaining popularity in fields such as human-computer interaction, augmented reality, and neuroergonomics, where understanding visual attention in naturalistic contexts is crucial.

The primary advantage of wearable eye trackers is their ability to capture eye movements in dynamic and ecologically valid settings. In contrast to remote eye trackers confined to controlled laboratory environments, wearable devices empower researchers to examine gaze behavior in everyday situations, including mobile device usage, driving, or outdoor activities. Additionally, wearable eye trackers afford participants increased freedom of movement, enabling them to interact with their environment while their eye movements are recorded.

However, wearable eye-tracking systems also pose certain challenges. These include relatively lower accuracy than stationary or remote systems, particularly in challenging environments with variable lighting conditions or user movement. Precise calibration of the glasses' position and orientation relative to the wearer's eyes is necessary. Comfort and usability can be additional concerns, as wearing eye-tracking glasses for extended periods may cause discomfort, potentially impacting data quality and participant compliance. Moreover, these devices rely on batteries, which may have limited lifespans and require frequent recharging or replacement. Furthermore, analyzing eye-tracking data from wearable devices can be more intricate due to factors such as head movements, motion artifacts from devices, and variations in gaze behavior across different real-world contexts.

Nevertheless, advancements in wearable technology and signal-processing algorithms continue to enhance these systems' accuracy, reliability, and usability for a broad spectrum of research applications. Table~\ref{table:eye-trackers-comparison} shows the comparison between remote and wearable eye trackers. 
\begin{table}[ht]
\caption{Comparison of Remote and Wearable Eye-Trackers}
\label{table:eye-trackers-comparison}
\centering
\begin{tabular}{p{0.13\linewidth} p{0.4\linewidth}p{0.4\linewidth}}
\toprule
\textbf{Feature} & \textbf{Remote Eye Tracker} & \textbf{Wearable Eye Tracker} \\ \hline
\textbf{Description} & Non-invasive systems mounted near a screen to monitor eye movements from a distance. & Lightweight glasses or head-mounted devices that record eye movements in real-world environments. \\ \midrule
\textbf{Manufacturers} & Example: Tobii, EyeLink, SMI & Example: Tobii, Pupil Labs, EyeLink, SMI \\ \midrule
\textbf{Components} & Incorporates a monitor or screen for stimulus presentation and infrared cameras positioned to detect reflections of infrared light from the eyes. & Integrates miniature cameras and infrared sensors into glasses or head-mounted devices. Includes both eye and scene cameras. \\ \midrule
\textbf{Use Cases} & Commonly used in laboratory settings for recording eye movements with a comfortable seating arrangement, e.g., medical research~\citep{remotemedical18} and educational research such as reading comprehension~\citep{abundis2018reading}. & Used in real-world environments, suitable for fields such as human-computer interaction, virtual and augmented reality, and neuroergonomics, e.g., market research~\citep{wedel2017review}, eye-gaze-based interaction in virtual reality~\citep{vrgazedinter17}, and driving~\citep{driving2015ap}. \\ \midrule
\textbf{Advantages} & High accuracy and precision, user-friendly, adaptable to various experimental setups, suitable for a broad spectrum of research applications, non-invasive. & Captures eye movements in dynamic and ecologically valid settings, offers mobility, allows for the study of gaze behavior in everyday situations, provides freedom of movement. \\ \midrule
\textbf{Limitations} & May have difficulty capturing eye movements during dynamic activities or in real-world scenarios, requires relatively fixed head position, susceptible to interference from occlusions. & May have lower accuracy in variable lighting conditions or with user movement, requires precise calibration, potential discomfort when worn for extended periods, relies on battery power, and data analysis can be intricate due to additional factors like head movements. \\ \midrule
\textbf{Integration} & Seamlessly integrates with existing computer setups and software. & Commonly integrated into glasses or head-mounted displays. \\ \midrule
\textbf{Nature} & Non-invasive and minimizes participant discomfort, allowing for extended experimental sessions. & Affords participants increased freedom of movement, enabling interaction with the environment while recording eye movements. \\ \midrule
\textbf{Challenges} & Capturing eye movements during dynamic activities or real-world scenarios, maintaining a fixed head position, dealing with occlusions. & Lower accuracy in challenging environments, precise calibration needed, potential discomfort for extended use, battery life limitations, intricate data analysis due to head movements and motion artifacts. \\ \midrule
\textbf{Technological Advances} & Advancements in screen-based eye trackers have led to greater spatial accuracy and temporal resolution, allowing for precise tracking under varying lighting conditions and at higher sampling rates. & Advancements in wearable technology and signal-processing algorithms continue to improve accuracy, reliability, and usability for research applications. \\ \bottomrule
\end{tabular}
\end{table}

\section{Calibration}
This section describes the process of calibrating different eye-tracking systems, an essential step for ensuring the accuracy and reliability of the collected eye-tracking data.

Eye-tracking calibration is the process of estimating the geometric characteristics of a participant's eyes as the basis for a fully customized and accurate gaze point calculation~\citep{ra2006calibration}. Given the uniqueness of each participant's eyes in shape, size, and movement, eye-tracking devices require calibration to effectively accommodate these individual differences. In addition, environmental factors such as lighting conditions, the position of the person's head, and distance from the screen to the eyes can affect the accuracy of eye-tracking data. Therefore, calibration is indispensable before collecting eye-tracking data to accurately determine the gaze point on a screen or within the environment~\citep{nystrom2013influence}.

A calibration procedure aims at establishing an accurate mapping between measured eye movements and the corresponding points of gaze on a screen or within an environment. In particular, eye tracker systems capture the user's eye movements as they sequentially fixate on predefined calibration points displayed on a screen or within an environment. By analyzing the relationship between these recorded eye movements and the known positions of the calibration points, the system estimates the transformation needed to accurately map eye movements to gaze positions. 
One commonly employed approach involves standard calibration, which utilizes linear or second-order models to estimate the relationship between eye movements and gaze positions based on the recorded data from fixations on calibration points~\citep{liu2018gazecali, morimoto1999frame}. 
Another prevalent technique is 2D mapping with interpolation~\citep{sheela2011mapping}, where users fixate on a grid of calibration points distributed across the screen or environment, and interpolation algorithms are used to estimate gaze positions between these points. Through these calibration processes, eye-tracking devices establish accurate mappings between eye movements and gaze positions. 

In the actual calibration procedure with eye-tracking devices, participants are typically instructed to direct their gaze towards a series of points that cover the area where relevant stimuli are presented, either on a screen or within the environment. These points are commonly arranged in a grid or pattern that comprehensively covers the viewing area. Throughout this process, the eye-tracking device continuously records the position of the user's eyes as they focus on each calibration point in succession. The participant may be prompted to blink or briefly look away between each calibration point to ensure precise tracking by the device. During the calibration process, the eye-tracking device collects data about the participant's gaze position and compares it to the known locations of the calibration points. This data is used to generate the mapping or calibration profile that allows the device to accurately track the participant's gaze during the data collection. Subsequently, we provide a more detailed description of two distinct types of calibration employed in eye tracking.

\begin{figure}[b]
  \centering
   \subfigure[Five-point screen-based eye-tracking calibration \& four-point validation in Tobii Pro Lab.]{{\includegraphics[width=0.42  \linewidth,keepaspectratio]{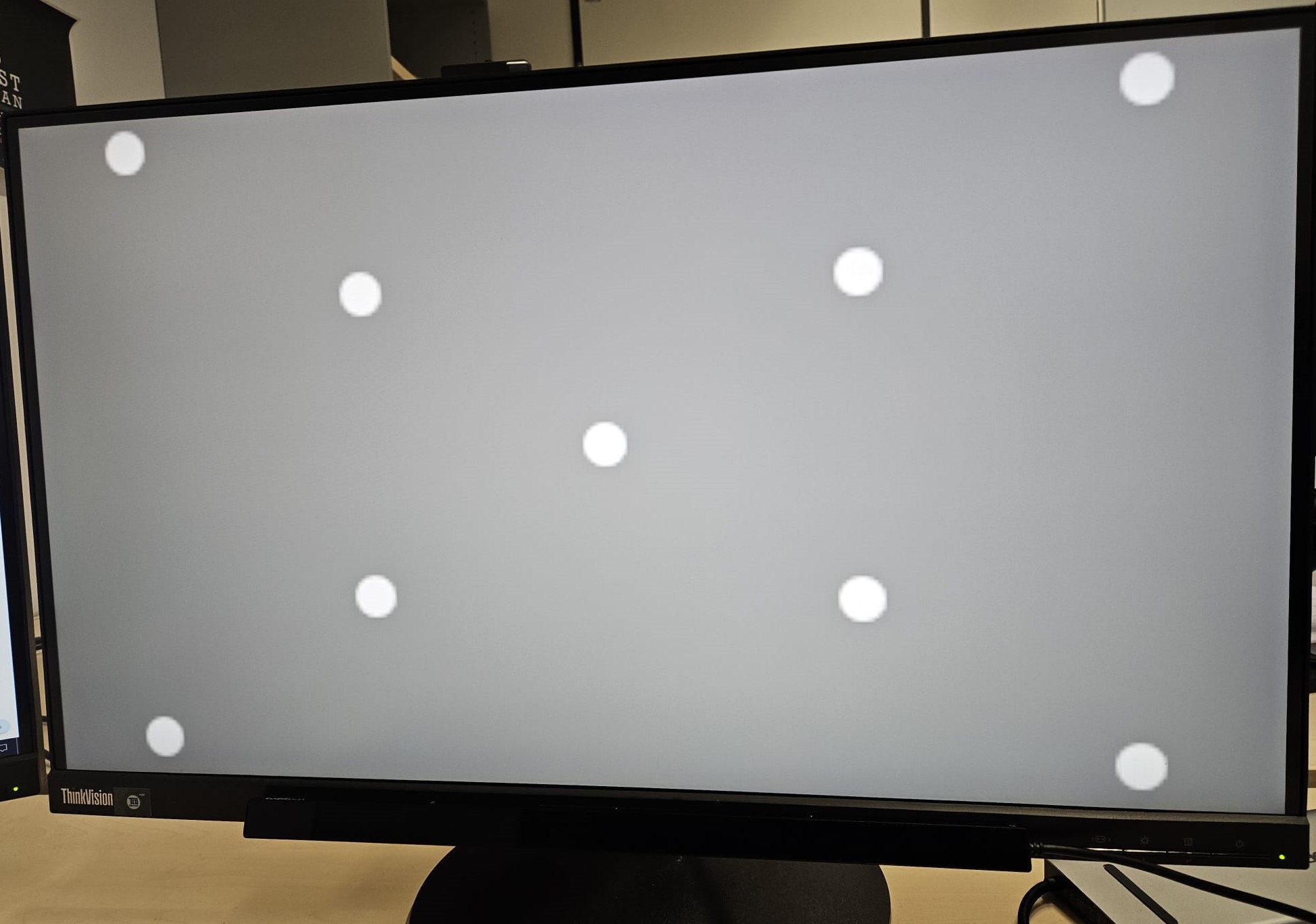}}}%
   \quad
   \subfigure[One-point wearable eye-tracking calibration in the Glasses 3 Controller Software.]{{\includegraphics[width=0.48\linewidth,keepaspectratio]{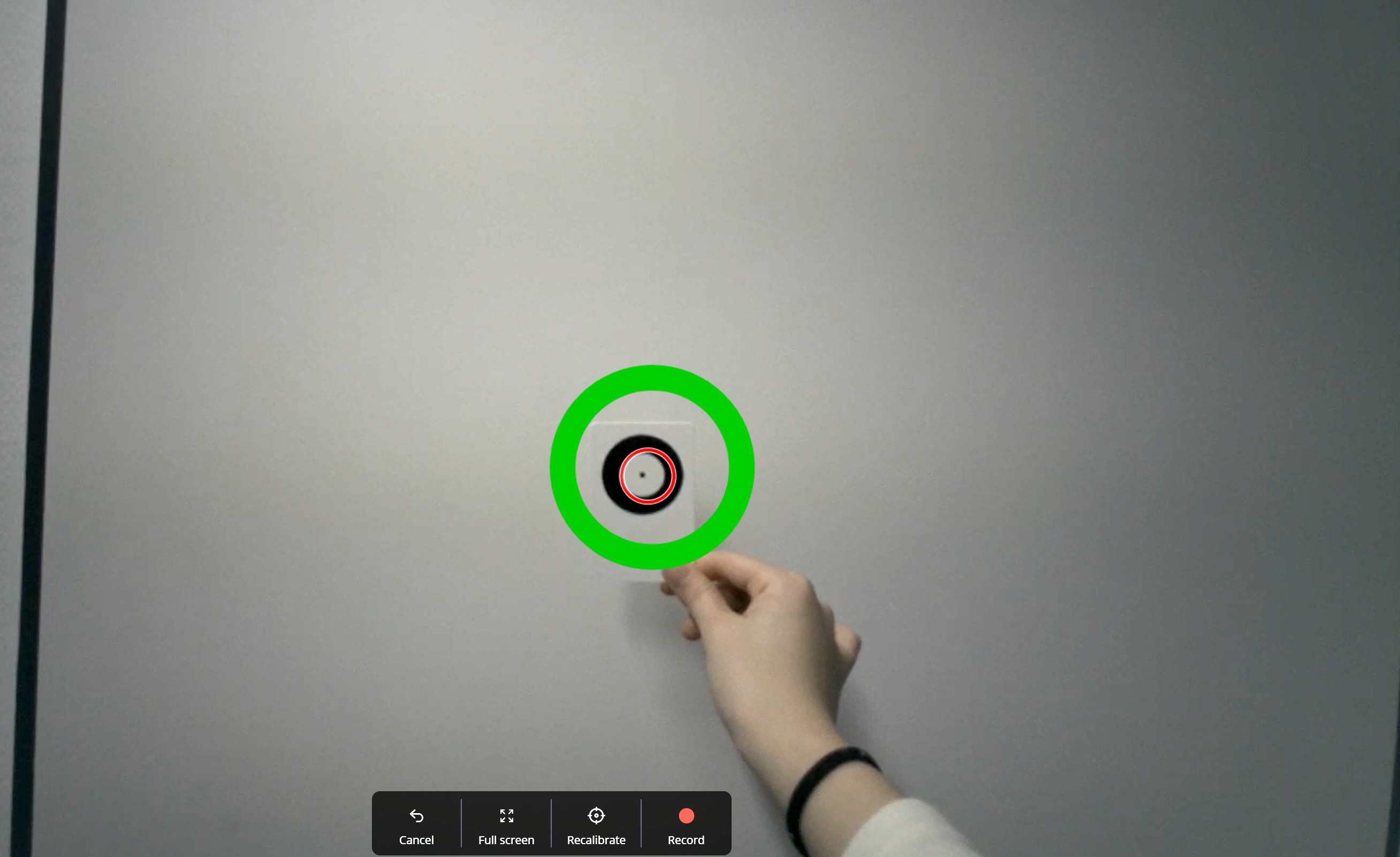} }}%
  \caption{Eye-tracking calibration.}
   \label{fig:cali_combine}
\end{figure}

\subsection{Screen-based Calibration}
The screen-based calibration method projects calibration points onto the 2D surface of the display monitor in a random sequence, a common process for calibrating remote eye-tracking devices. The participant is then instructed to fixate on each calibration point sequentially. The number of calibration points used in this process commonly ranges from 2 to 16 \citep{holmqvist2011eye, cali2018}, with 9-point calibration being a widely adopted standard~\citep{tobii_pro_calibration}. One of the primary motivations behind screen-based calibration is the inherent variability in digital displays, including differences in size, resolution, and aspect ratio. These variations can significantly impact the accuracy of eye-tracking data if not properly calibrated. The eye-tracking system can account for these display discrepancies through the calibration process, ensuring precise and reliable gaze tracking across different display configurations. Screen-based calibration is important because digital displays can vary in size, resolution, and aspect ratio, affecting eye-tracking data quality.

An illustration of a screen-based 5-point calibration and 4-point validation is depicted in Figure~\ref{fig:cali_combine} (a). The 4-point validation, which follows the calibration process, involves presenting four points at new locations to verify the eye tracker's accuracy in tracking the participant's gaze direction. Successful calibration is confirmed if the gaze data aligns with the predetermined target positions; otherwise, recalibration is required \citep{tobii_validation}.

\subsection{Wearable Calibration}
Wearable calibration is primarily employed in scenarios where a participant's eye movements must be tracked in real-world environments rather than on traditional computer monitors or digital displays. This calibration method is commonly associated with wearable eye-tracking devices, such as eye-tracking glasses and head-mounted displays. These wearable eye-tracking devices may use various calibration procedures tailored to the device specifications and the specific application requirements~\citep{holmqvist2011eye, santini2017calib}. For instance, when using eye-tracking glasses, participants are instructed to direct their gaze toward calibration targets positioned in the real environment. These targets may be presented on a specially designed calibration card displaying the points. Two commonly employed calibration procedures in wearable eye tracking are the single-point and multi-point calibration methods~\citep{santini2017calib}. In single-point calibration, participants focus their gaze on a single calibration target, while in multi-point calibration, they fixate on multiple calibration targets distributed across the environment. Figure~\ref{fig:cali_combine} (b) depicts an example of a calibration card used for wearable eye-tracking glasses. 

\subsection{Slippage Compensation}
Slippage refers to the unintended movement of the eye-tracking device on the participant’s head during a recording session. This issue is frequently encountered during long recordings or when participants are free to move~\cite{santini2018art,niehorster2020impact}. In the recording sessions, participants may push the device up on their nose, remove it momentarily to adjust their hair or rub their eyes, or even move it through normal facial movements like speaking or smiling. To address this issue, manufacturers have included stability features, such as tighter headbands or additional securing mechanisms, to prevent the device from moving out of place. Despite these efforts, slippage remains a common challenge, often requiring researchers to take additional steps to ensure data accuracy. 

Several techniques have been proposed in the literature to reduce or compensate for slippage. These include the classical pupil-glint vector technique~\cite{kolakowski2006compensating} and determining camera translation, which may involve using eyelid templates~\cite{karmali2004automatic}, eye corner tracking~\cite{pires2013visible} and monitoring differences in gain values~\cite{kolakowski2006compensating}. Additionally, saliency maps are useful for detecting and correcting shifts~\cite{sugano2015self}, thereby adjusting for unintended movements. Techniques such as fast and unsupervised calibration~\cite{santini2017calibme}, recalibration~\cite{lander2016time}, and auto-calibration schemes~\cite{huang2016building} can also mitigate the effects of slippage.

In addition to these methods, using an eye tracker and eye tracker accessories designed to manage slippage could be a more efficient and feasible option. Eye trackers often come equipped with accessories like head straps, cloth clips, helmets, or custom facial plastic molds, all of which help to secure the device firmly on the user during recordings. Head straps are elastic or adjustable bands that wrap around the head to effectively stabilize the eye tracker. Commonly used in virtual reality headsets and other head-mounted devices, head straps provide a reliable attachment method. Cloth clips, which can be attached to hats, glasses, or other headgear, are less intrusive than head straps and offer a quick way to attach and detach the eye tracker. Helmets could be particularly useful in applications requiring robust mounting of the eye-tracking device. Additionally, custom facial plastic molds, designed for an individual’s face, provide the most secure and precise fit possible. However, this method is less common due to the higher cost and complexity of creating personalized molds.

\subsection{Practical Advice}
In the calibration process, one of the most important factors is lighting. To achieve optimal accuracy and precision with wearable glasses, both direct sunlight and excessively dark settings should be avoided; moderate lighting is most suitable. Users need to ensure that the glasses are fitted comfortably during calibration, as adjustments made afterward can negatively affect the accuracy. Additionally, lighting must align with the conditions of the actual experimental environment.

Remote eye trackers, which often employ infrared lights, perform better in darker rooms. If users wear glasses, calibration accuracy can be significantly diminished. In such cases, a minor adjustment in the positioning of their glasses might enhance the fit and improve accuracy.

In the calibration process, a neutral and non-distracting background is better for both types of eye trackers. Intricate patterns or moving objects can interfere with eye-tracking accuracy during calibration. It is crucial to clearly explain the calibration process and subsequent steps to prevent users from moving their heads to ask questions after the calibration. In the end, calibration accuracy should be verified numerically (through accuracy and precision metrics) and visually, by ensuring that calibration points cluster closely around the target dot with minimal spread. If the calibration results are inadequate, the process should be repeated to ensure the experiment's validity.

\section{Data Collection}

The data collection procedure in eye-tracking studies is a systematic process that establishes the experimental setting, followed by experiment design and planning. This procedure concludes with the recruitment of participants and the collection of eye movement data. This chapter will delve into every step of the eye-tracking experiment including setup, piloting recording and troubleshooting, providing a comprehensive and hands-on guide for each aspect.

\subsection{Setup}
\label{subsec_setup}

\begin{figure}[b]
    \centering
    \includegraphics[width=.5\linewidth]{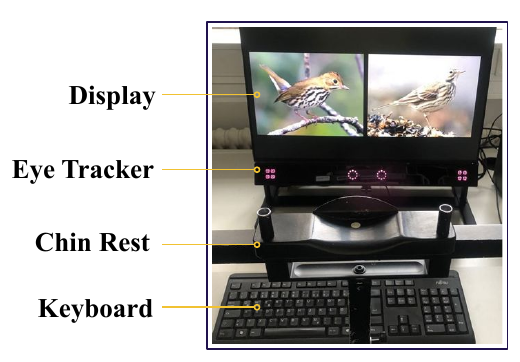} 
    \caption{Eye-tracking data collection setup using a remote eye tracker.}
    \label{fig:setup}
\end{figure}

The setup of the experimental environment plays a decisive role in the successful implementation of user studies. Figure~\ref{fig:setup} illustrates an exemplary setup for data collection using a remote eye tracker and a computer monitor to display the visual stimuli.
Eye trackers with a high sampling frequency can yield high-resolution data. Similarly, using monitors with higher refresh rates can positively impact data quality. In this regard, it is essential to consider the experimental objectives; for instance, features such as saccade latencies may require the use of 240Hz monitors. Consistency in the experimental setup is also crucial in eye-tracking research. To ensure the comparability of eye-tracking data across different participants, the eye-tracking devices and monitors used in the experiments should remain identical.
Additionally, the configuration of the desktop or laptop used in the experiment can affect the data collection process and should be carefully considered. In the experimental design, connecting eye trackers with additional extension cables or hubs may affect the stability of their connection. Therefore, it is advisable to avoid using additional extension cables for the eye trackers and instead connect them directly to the computer. If additional cables are used or if the devices are connected to ports on the monitor, thorough testing is recommended. Moreover, utilizing ports with higher speeds on the computer can enhance stability.

Positioning the eye trackers and configuring the monitors are crucial steps in the setup process. After positioning the eye tracker, configuration typically involves using dedicated software for eye trackers, which may require specific information about the monitor, such as its width, length, and resolution. 
Attaching the eye trackers directly to the screen simplifies the configuration process, requiring less detailed information about the eye tracker's position. 
However, if the eye tracker is positioned elsewhere, precise measurements of the angles and distances between the eye tracker and monitor are essential for accurate configuration. 
Eye tracker models equipped with their own screens require less detailed information for configuration due to their built-in screens. Once the eye tracker has been configured, it is important to refrain from moving or changing any setup components, as even minor adjustments can adversely affect the data collection process. In case of any changes, reconfiguration is necessary before proceeding with the experiment. This configuration process is essential, as the accuracy of the collected data hinges on providing precise and accurate information during setup.

Proper screen resolution adjustment is also essential, as eye-tracking data logs typically record gaze positions in centimeters or pixels. It is advisable to adhere to the resolutions recommended by the eye-tracking software being used. In cases where specific recommendations are unavailable, opting for higher resolutions may be beneficial, but it is important to ensure that the computer's processing power can adequately support them. Conversely, excessive computational load during the experiment can adversely affect the results, so lower resolutions can be used if necessary. The decision between higher and lower resolutions should be carefully weighed considering various factors. 

Careful design of the experimental environment is essential to mitigate external stimuli such as sound and light. Consistency in external factors, such as lighting conditions, is crucial if experiments are conducted at different times. If the eye-tracking device can function in low-light conditions~\citep{tobiiLight}, covering windows and turning off lights could provide an optimal experimental setup, as such external light sources can fluctuate throughout the day. When collecting data from multiple participants simultaneously, it is important to ensure that each participant is not affected by others' presence or actions. During the experiment, the experimental room's door can be closed, and a warning sign indicating that the experiment is in progress can be placed outside. 

\begin{figure}[t]
\centering
\includegraphics[width=0.3\linewidth]{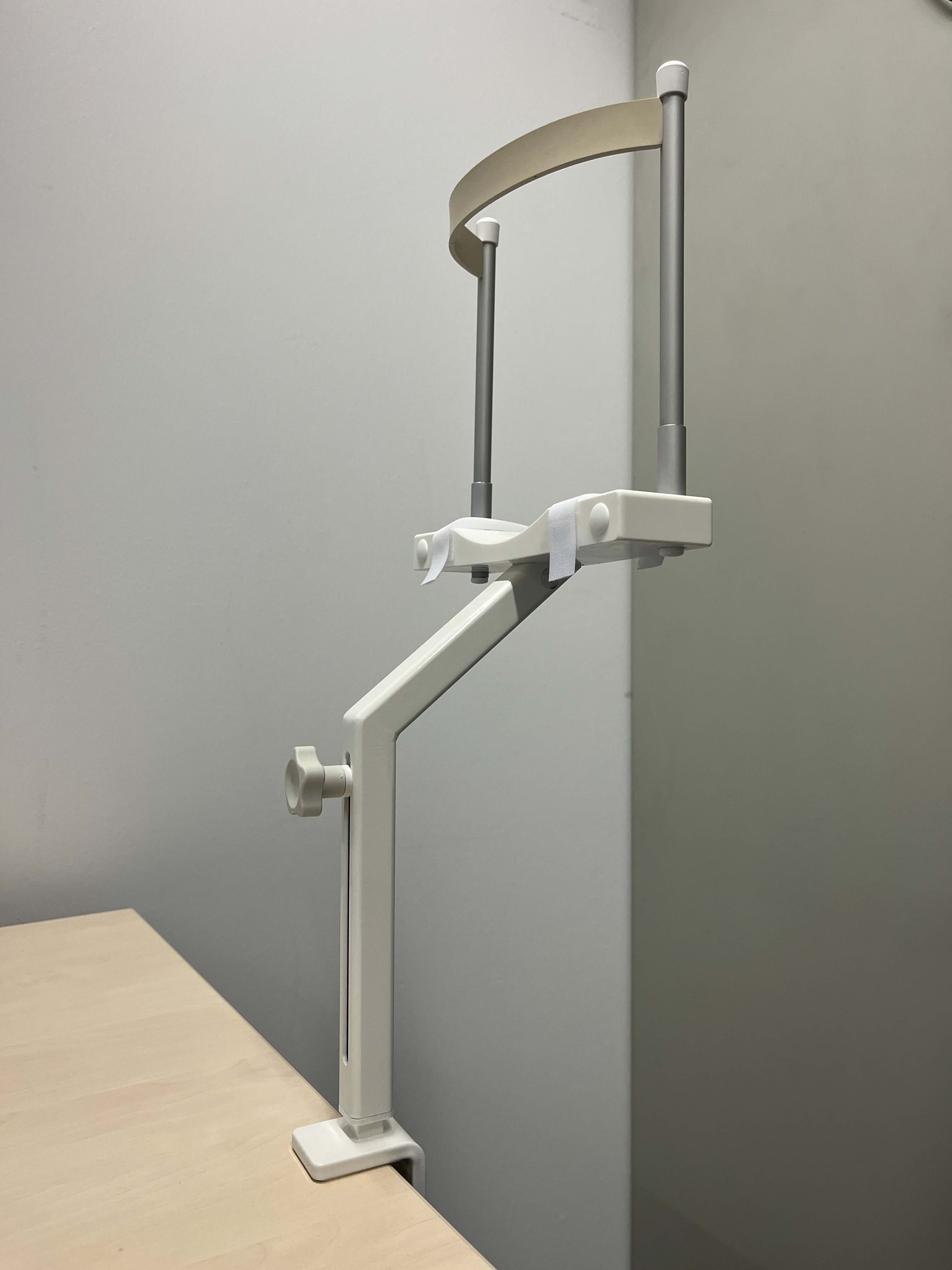}
\caption{Example of a chin rest used to stabilize the participant's head during experimental procedures.}
\label{fig:chinrest}
\end{figure}

Furthermore, experimenters may opt to use a chin rest as depicted in Figure \ref{fig:chinrest}, or other head-stabilizing methods during the experiments to improve eye-tracking data quality. However, such methods may cause discomfort in participants and create an artificial experimental setting. Thus, applying these methods necessitates carefully evaluating their advantages and disadvantages in balancing experimental control with ecological validity~\citep{MCGRATH1995152} preservation. Ecological validity refers to how accurately research findings represent real-world phenomena and their generalizability to naturalistic settings. 
A chin rest can effectively minimize head movements and maintain consistent positioning of participants throughout the experiment. This implementation enhances the accuracy and reliability of eye-tracking data while standardizing the experimental setup across all participants. However, restricting natural head movements can be problematic for experiments requiring participants to move their heads freely. Therefore, it is essential to carefully weigh the benefits of employing a chin rest against its potential limitations, particularly regarding data quality and ecological validity implications.

\subsection{Experiment Design}

In the previous Section \ref{subsec_setup}, we discussed the process of configuring hardware and establishing the experimental environment for data collection. In this section, we provide an overview of key considerations essential for the design of an eye-tracking experiment.

Before designing the eye-tracking experiment, careful consideration should be given to how stimuli are presented and what data needs to be recorded. Some eye-tracking devices may require the use of specific software for experiment preparation. When using videos or images as stimuli in the experiment, it is important to adjust their resolution carefully to ensure participant comfort and facilitate post-processing. It may be advisable to avoid displaying stimuli in full-screen mode and instead opt for a smaller area surrounded by a black frame. This approach helps to prevent less accurate recording of gaze data on the monitor's edges and minimize participant distraction. 

Participants' gaze behavior is typically recorded using eye-tracking software. Still, in certain situations, recording keyboard and mouse movements or voice inputs may be necessary to capture participant responses or answers to specific tasks during the experiment. If the software does not provide these recordings, additional scripts should be integrated to capture those data. In particular, it is crucial to ensure that timestamps are included in the recordings; however, synchronizing these timestamps between the recording and the eye tracker logs can be challenging. This challenge may be further complicated by using different time options, such as computer time and eye tracker recording time. Consequently, checking these timestamps and analyzing any constant or varying bias is essential, as they can later pose problems during the data post-processing. If complete synchronization of the timestamps is not feasible, one potential solution is to include synchronization logs in the scripts or redesign the experiment by incorporating press-button instructions following each stimulus.

In addition to details on the experimental environment and equipment, designing the experiment itself is another essential step for obtaining valid findings. In eye-tracking studies, two common experimental designs are typically employed: the between-subjects design and the within-subjects design. These designs differ primarily in how the experimental conditions are assigned to the participants. In a between-subjects design, each participant is exposed to only one experimental condition, with participants randomly assigned to these conditions to ensure each group experiences a different condition. Conversely, a within-subjects design involves the same participants being exposed to all experimental conditions. One important factor that can affect the reliability of experimental results is the order effect, which refers to the influence of the stimulus presentation sequence on a participant's responses. The order in which stimuli are presented can significantly impact the participant's response and perception, potentially leading to biased results. To mitigate this effect, researchers often randomize the presentation order of stimuli, minimizing potential confounding factors and ensuring the objectivity and reliability of the experimental results. Additionally, in a within-subjects design, it is important to consider the potential impact of the training effect, which refers to the improvement in performance that occurs as a result of repeating the same or similar tasks over time. In addition to potential performance increases in certain tasks, familiarity with the tasks or conditions may lead to decreased participant engagement, potentially leading to a loss of interest or focus. 

Careful consideration of each stimulus's duration and presentation timing is essential in optimizing the experimental design and mitigating potential confounding factors such as participant fatigue. Participant fatigue can significantly affect eye-tracking data quality and validity over time, reducing attention and slower ocular responses. Therefore, avoiding excessively prolonged sessions and providing adequate rest periods to prevent fatigue is imperative.
Throughout the experiment, changes in the participant's posture or disruptions in the eye-tracking signal may necessitate the recalibration of eye trackers. Particularly in long experiments, incorporating breaks that include recalibration may help maintain consistency in the collected data and ensure that the results accurately reflect the participant's gaze behavior. 

During the data collection process, it is important to acknowledge that participants may not be familiar with the equipment and experiment. Merely providing instructions alone may not be sufficient for their full understanding. Integrating an acclimation phase before the actual experiment is highly recommended to enhance understanding. This phase should consist of a series of specially designed stimuli, different from but related to those used in the main experiment, to familiarize participants with the tasks. Encouraging participants to ask questions during this period is important for clarifying uncertainties and improving their comfort level for the experiment. These may increase the quality and reliability of the data collected by minimizing the disruption during the experiment.

Moreover, it is essential to establish the budget during the experimental design phase, considering the required number of participants. Compensation should be provided based on the duration of the experiment to encourage participant involvement and comply with legal requirements. It is essential to note that the compensation rate should be determined in accordance with the minimum hourly rate applicable in the region where the experiment is to be conducted. 

\subsection{Pilot Study}

Piloting is pivotal in eye-tracking experiments, serving as a critical preparatory phase before the main study. By conducting a smaller-scale version of the experiment with a limited number of participants, researchers can identify and address potential issues or inefficiencies in the design. The piloting process allows for the refinement of the experimental design, visual stimuli used, and experimental procedures to ensure they are suitable for obtaining the research objectives. Pilot studies also enable researchers to pinpoint potential technical issues with the eye-tracking equipment, software, or other components of the experiment, therefore preventing data loss or errors during the main study. Moreover, they provide an opportunity to assess participants' comprehension of instructions and the suitability of tasks, facilitating adjustments to task difficulty or duration as needed. Additionally, pilot studies aid in estimating the time required for data collection per participant more accurately, enabling efficient planning and scheduling for the main experiment. This, in turn, facilitates a more precise estimation of the budget required for the main experiment. Furthermore, pilot studies allow for assessing the quality of the eye-tracking data collected, including accuracy, precision, and sampling rate. This assessment informs researchers of any necessary adjustments to improve data quality and ensures that data analysis procedures are robust and effective. If issues with data processing or missing information arise, researchers can refine the experimental procedure to obtain additional measurements, thus enhancing the overall reliability and validity of the study.

Notably, controlling the experimental environment is often challenging for studies involving wearable eye trackers as they are conducted in real-world settings. This lack of control introduces variables such as lighting conditions, which can lead to distractions and significantly impact the quality of visual recordings. Therefore, during the pilot study, it is also essential to experiment with different lighting conditions and analyze the resulting data to verify its quality.

\subsection{Experimental Procedure}
\begin{figure}[b]
\centering
\includegraphics[width=1\linewidth]{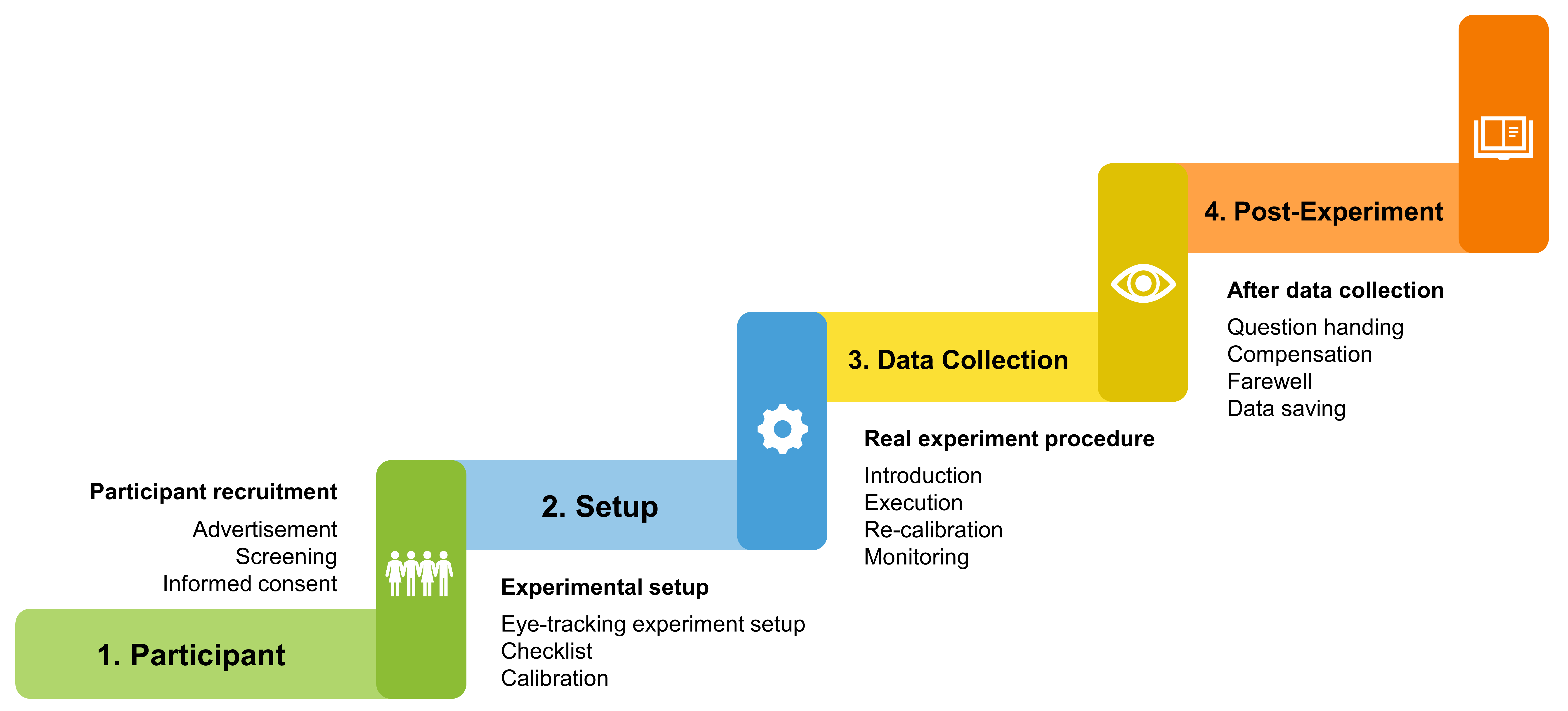}
\caption{Experimental procedure diagram.}
\label{fig:exp_procedure}
\end{figure}

As depicted in Figure~\ref{fig:exp_procedure}, the initial step in conducting eye-tracking experiments entails recruiting participants, which requires efforts to draw a representative sample of participants from the target population. 
A comprehensive advertisement is initially constructed, outlining the experiment's nature, time commitments, and any associated compensation. These advertisements should provide only general information about the study, avoiding extensive details about the experiment's purpose and methodology to maintain experimental integrity while offering sufficient information to potential participants.
The advertisement can be disseminated across various platforms, including institutional bulletin boards, online forums, and social media, to reach a large and diverse pool of potential participants. Additionally, student groups on platforms such as WhatsApp, Slack, Telegram, and university email groups can serve as useful alternatives for participant recruitment. Alongside the advertisement, brief questionnaires can be utilized to gather basic demographic details such as age and language proficiency in predetermining participant eligibility based on broad inclusion and exclusion criteria. 

Following the recruitment phase, screening involves a more detailed assessment of selected participants to ensure they meet the specific criteria for the eye-tracking experiment. Questionnaires can be used to gather information based on the specific experiment's criteria, including inquiries related to visual acuity requirements due to the nature of the eye-tracking technology.
Participants who pass the screening process can participate in the experiment. Before taking part in the experiment, participants are provided with an informed consent form. This document outlines the study's nature, procedures, potential risks and benefits, and measures taken to ensure participant confidentiality. Upon approval and signing of the consent form, participants can proceed to the experiment, be added to the participant pool, and participate in the experiment. 

Preparation of the eye-tracking experiment setup should be completed before the participant arrives. The experimenter should ensure that all necessary components are properly connected and operational and adjust the eye tracker configuration as needed. Mistakes and delays can be minimized by completing these preparations in advance, thus ensuring a legitimate data collection process. Using a checklist (see Table~\ref{table:eye-tracking-checklist}) for experiment preparation may be essential to streamline setup procedures and ensure all necessary steps are executed.

\begin{table}[t]
\caption{Checklist for Conducting Successful Eye-Tracking Experiments}
\centering
{\renewcommand{\arraystretch}{1.39} 
\begin{tabular}{p{0.13\linewidth} p{0.7\linewidth}p{0.03\linewidth}}
\toprule
\textbf{Category} & \textbf{Checklist Item} \\
\midrule
\textbf{Initial} & - Use eye trackers with a high sampling frequency for accurate data. & \scalebox{1.8}{\Square} \\
\textbf{Setup}                           & - Choose monitors that have high refresh rates. & \scalebox{1.8}{\Square} \\
                              & - Ensure eye trackers are connected directly to the computer for reliability. & \scalebox{1.8}{\Square} \\
                              & - Position and configure the eye tracker accurately. & \scalebox{1.8}{\Square} \\
                              & - Adjust screen resolution appropriately. & \scalebox{1.8}{\Square} \\
                              & - Control external stimuli such as sound and light to prevent distractions. & \scalebox{1.8}{\Square} \\
                              & - Utilize head stabilizers (e.g., a chin rest) for data quality. & \scalebox{1.8}{\Square} \\
\midrule
\textbf{Experiment}   & - Select an appropriate experimental design (between-subjects or within-subjects). & \scalebox{1.8}{\Square} \\
 \textbf{Design}                          & - Determine the types of stimuli (e.g., images, videos, text) to be presented. & \scalebox{1.8}{\Square} \\
                           & - Ensure the resolution of the stimuli is appropriate for your setup. & \scalebox{1.8}{\Square} \\
                           & - Determine additional interactions (keyboard, mouse) with proper timestamps. & \scalebox{1.8}{\Square} \\
                           & - Manage session duration to avoid fatigue. & \scalebox{1.8}{\Square} \\
                           & - Recalibrate during breaks for long sessions to maintain data quality. & \scalebox{1.8}{\Square} \\
\midrule
\textbf{Pilot} & - Conduct to refine the experimental design and procedures. & \scalebox{1.8}{\Square} \\
\textbf{Study}                   & - Assess participant instruction comprehension. & \scalebox{1.8}{\Square} \\
                     & - Estimate the time required for data collection. & \scalebox{1.8}{\Square} \\
                     & - Verify eye-tracking data quality. & \scalebox{1.8}{\Square} \\
                     & - Validate and test your analysis method on the collected data. & \scalebox{1.8}{\Square} \\

\midrule
\textbf{Experimental} & - Recruit and screen participants. & \scalebox{1.8}{\Square} \\
\textbf{Procedure}                              & - Provide informed consent forms. & \scalebox{1.8}{\Square} \\
                                & - Prepare and test the eye-tracking setup in advance. & \scalebox{1.8}{\Square} \\
                                & - Explain the experiment procedure to participants. & \scalebox{1.8}{\Square} \\
                                & - Ensure participants understand the experiment procedure. & \scalebox{1.8}{\Square} \\
                                & - Ensure that participants are comfortably seated. & \scalebox{1.8}{\Square} \\
                                & - Accurately calibrate the eye-tracking system. & \scalebox{1.8}{\Square} \\
                                & - Maintain consistent participant position. & \scalebox{1.8}{\Square} \\
                                & - Provide compensation and ensure a positive experience. & \scalebox{1.8}{\Square} \\
                                & - Save and store data securely. & \scalebox{1.8}{\Square} \\
\midrule
\textbf{Troubleshooting} & - Restart software or computer if needed. & \scalebox{1.8}{\Square}\\
                         & - Adjust the eye tracker and screen position or angle for better capture. & \scalebox{1.8}{\Square} \\
                         & - Ensure accurate calibration, repeat if necessary. & \scalebox{1.8}{\Square} \\
                         & - Offer short breaks before recalibration attempts. & \scalebox{1.8}{\Square} \\
\bottomrule
\end{tabular}
}
\label{table:eye-tracking-checklist}
\end{table}

Upon the participants' arrival, the experimenters should welcome them warmly and respectfully, creating an atmosphere of comfort and open communication. Participants should be provided with a simple and non-technical explanation of how the eye tracker works. It is essential to ensure that the participant is comfortably seated in front of the eye tracker, with adjustments to accommodate their posture in case the experiment utilizes a remote eye tracker. Maintaining a consistent position throughout the experiment is essential to preserving the quality of the eye-tracking data. If supplementary data from devices such as a keyboard or mouse are being collected, the experimenters should ensure that these devices are readily accessible to the participants to prevent any posture and position changes during the experiment.

Experimenters should clearly provide explanations of the experimental procedure. Once the experimenters are confident that the participant understands the experimental procedure, they can proceed to the experiment. The participant's position should be appropriately adjusted, and the eye-tracking system should be successfully calibrated. This calibration process may need to be repeated until small gaze errors indicating accurate tracking are achieved. Throughout the experiment, the experimenter should be available to ensure that the participant is comfortable and that the eye-tracking data is being properly recorded. Additionally, the experimenter must ensure the calibration process is performed with minimal gaze errors in case of a break in the experiment. 

After completing the eye-tracking experiment, participants can ask any questions they may have about the experiment. Experimenters should respond to these inquiries openly, clearly, and respectfully. If compensation is applicable, participants should receive it at the end of the session as agreed upon. Lastly, the experimenter should make sure a respectful and friendly farewell happens, ensuring they leave with a positive perception of their participation. After the participant leaves, all logs and experimental data must be saved and securely stored for future processing and analysis.

\subsection{Troubleshooting}

Occasionally, issues may arise with the eye-tracking software or computers during an experiment. Restarting the software might offer a quick resolution when faced with such challenges. If this proves ineffective, rebooting the computer and reconnecting the eye-tracking device could also resolve the issue. In situations where the eye trackers cannot capture one or both eyes, external interference from nearby objects may be a reason. Adjusting the eye tracker's position or angle could help solve the problem. If the problem persists, attempting to restart the eye tracker, reconfiguring it, and verifying its functionality may be necessary. 

One of the most common problems encountered in eye-tracking experiments is the inability to achieve accurate and precise calibration, even when the eyes are properly captured. The experimenter needs to ensure that the participant understands the calibration instructions and gazes directly at the calibration points as they appear. This issue is especially prevalent among participants wearing glasses. Repeating the calibration process and potentially adjusting the participants' position may help resolve the problem. In such situations, providing the participant with a short break of 2-3 minutes before attempting calibration again could be beneficial.

\section{Eye-tracking Data Processing Pipeline}

\begin{figure}[t]
\centering
\includegraphics[width=0.75\linewidth]{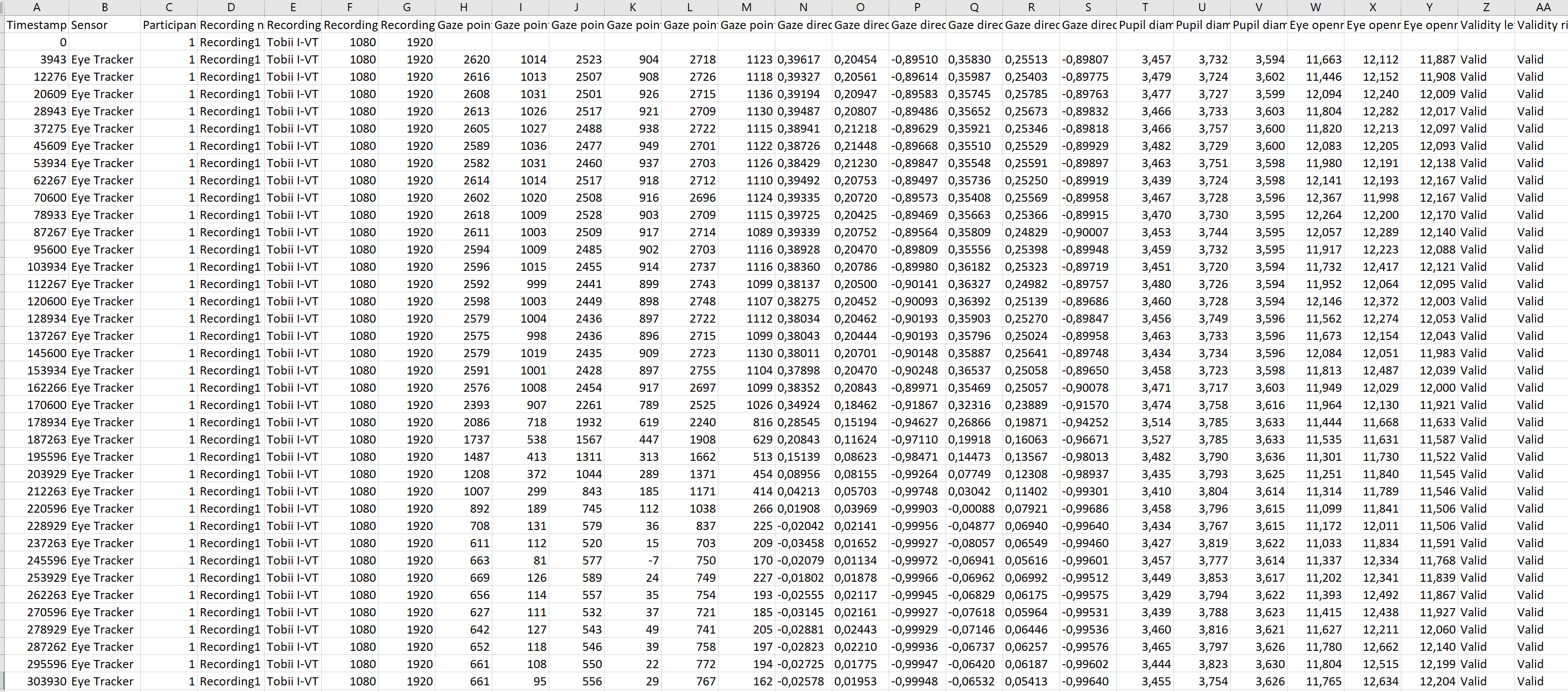}
\caption{An example of eye-tracking data collected from a Tobii remote eye tracker, formatted in CSV.}
\label{fig:csvdata}
\end{figure}

After data collection, the next step involves processing the raw eye-tracking data acquired from participants to prepare it for further analysis. The raw eye-tracking data retrieved from eye trackers is typically structured in tabular formats. Such datasets encompass multiple columns representing diverse parameters recorded by the eye-tracking system. Each row in the dataset corresponds to a specific time point or event during the experiment, with columns capturing data such as timestamp, eye status, gaze coordinates (x, y, z in 3D context), pupil size, and any additional experimental variables or annotations. Raw eye-tracking data can be stored or exported from eye-tracking software in various formats, including CSV (Comma-Separated Values), TSV (Tab-Separated Values), or proprietary file formats specific to the eye-tracking software or hardware used in the experiment. Figure~\ref{fig:csvdata} presents an example of eye-tracking data collected from a Tobii remote eye tracker formatted in CSV.

This section outlines a comprehensive data processing pipeline for readers to follow in their own user studies. The pipeline primarily encompasses data cleaning, data filtering, and feature extraction.

\subsection{Raw Data Cleanup} 
The first step of data processing involves excluding invalid raw data collected from participants. This process consists of three aspects, detailed in the following.   

\textbf{Excluding incorrectly calibrated data.} Data collected from participants with inaccurate calibration should be excluded from further analysis. Incorrect calibration can lead to inaccuracies in the recorded eye movements, compromising the reliability of the data. Therefore, it is essential to identify and remove such data to ensure the validity of subsequent analyses. In typical eye-tracking experiments, researchers ensure proper calibration before data collection. However, despite efforts to achieve accurate calibration, there are instances where calibration may not be successful or may degrade over time without immediate detection. For example, the calibration process may not yield accurate results despite repeated attempts initially. This could be due to factors that could not be solved, such as participant discomfort, eye illness, poor calibration techniques, or equipment malfunctions. In such cases, experimenters should proceed with the experiment rather than make further attempts to avoid causing frustration for participants. Additionally, even if successful calibration is achieved and participants are instructed to maintain a stable head position (in remote eye tracking) or keep the eye tracker stable on their head (in wearable eye tracking), if participants fail to adhere to these criteria during the experiment, the experimenters should not terminate the experiment but instead mark the data for potential exclusion in subsequent analysis.

\begin{figure}[t]
\centering
\includegraphics[width=0.75\linewidth]{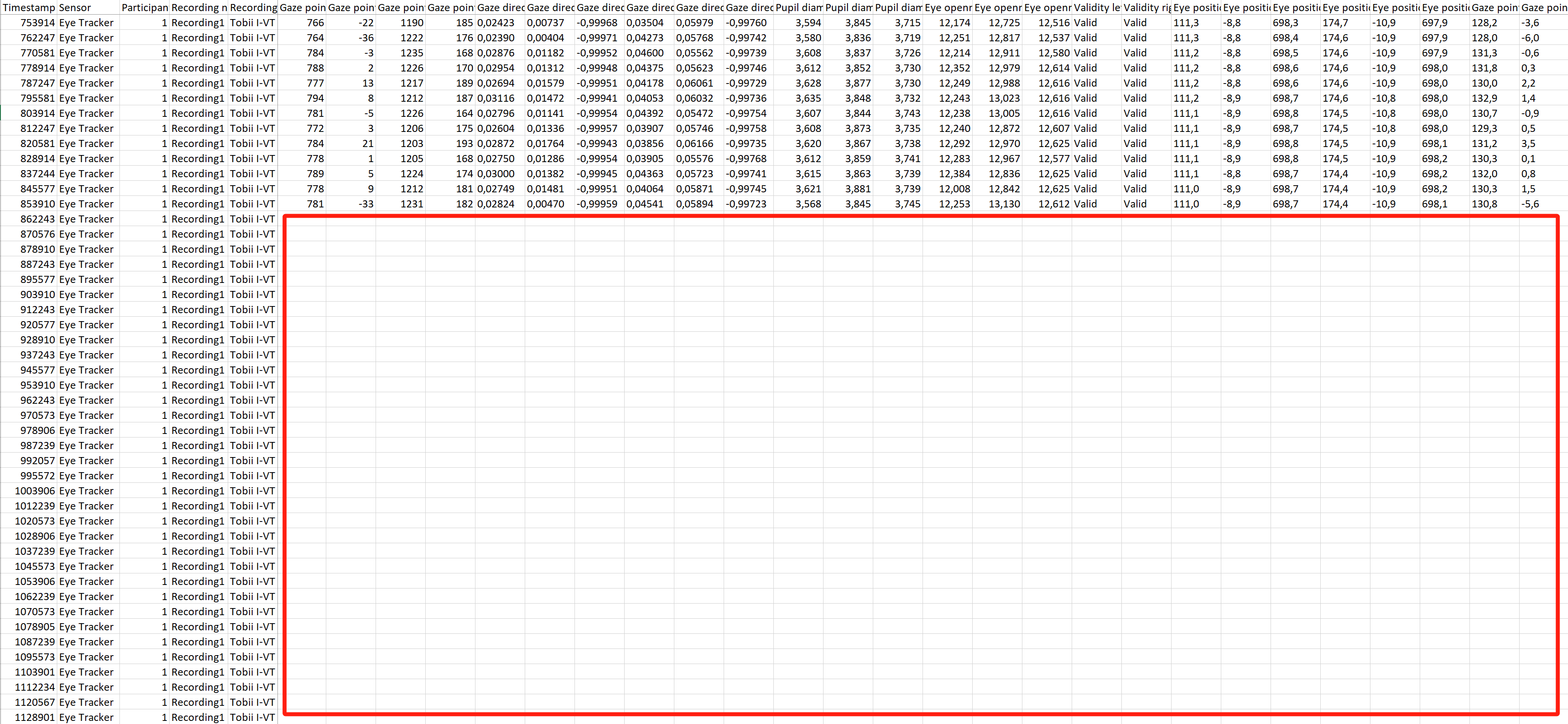}
\caption{An example of invalid eye-tracking data collected from a Tobii remote eye tracker, formatted in CSV. The rectangular region highlights the missing eye-tracking signals.}
\label{fig:csvmissing}
\end{figure}

\begin{figure}[t]
\centering
\includegraphics[width=1\linewidth]{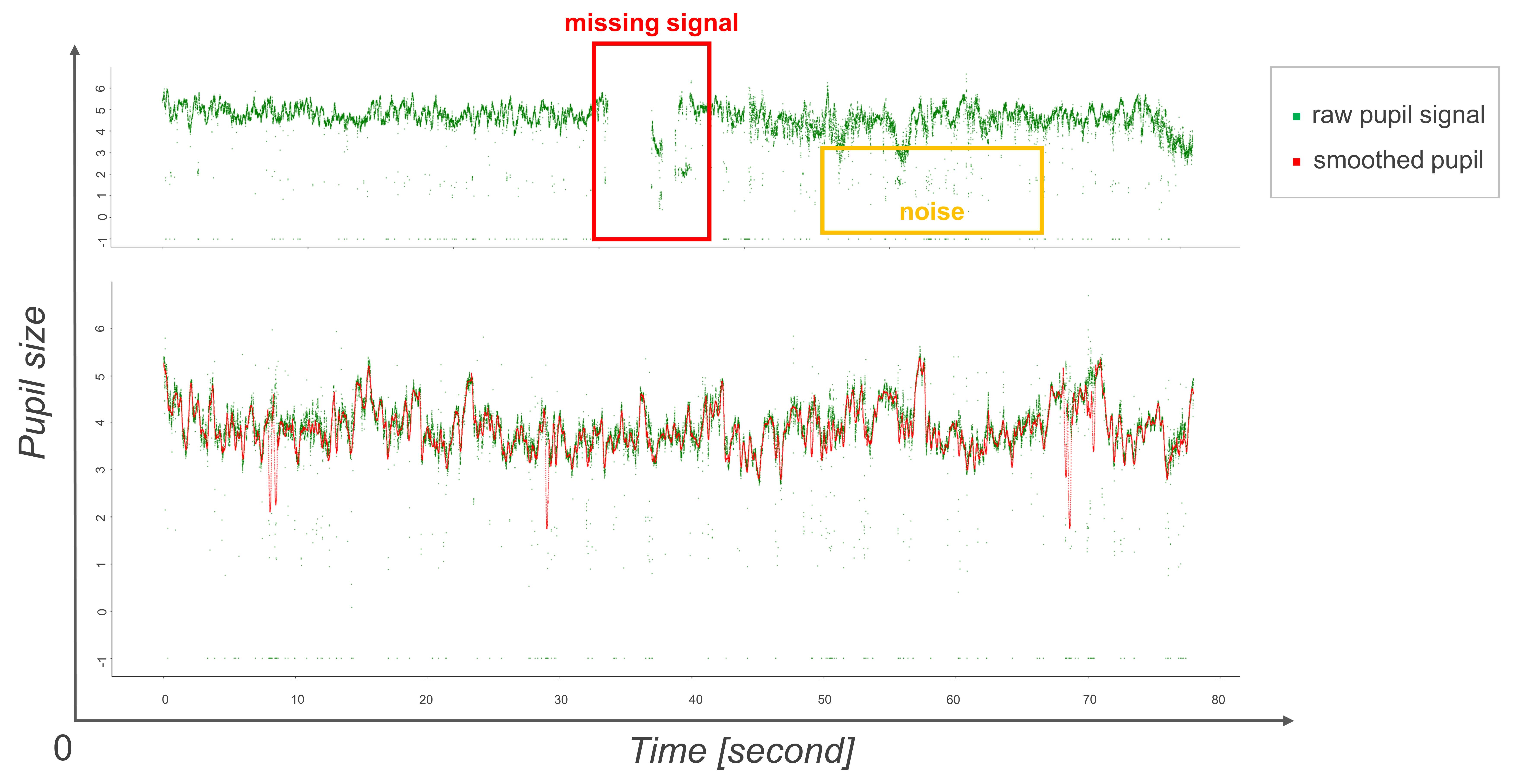}
\caption{An example of eye-tracking pupil signals collected from a Tobii remote eye tracker. The rectangular region highlights the missing pupil signals; the yellow rectangular region highlights the noise; the bottom figure displays the smoothed and interpolated pupil signals.}
\label{fig:pupilsignal}
\end{figure}

\textbf{Excluding incomplete data.} Incomplete data sets should also be excluded from analysis, which includes two scenarios. First, data collection was terminated during the experiment, either due to passive termination caused by hardware issues (e.g., eye trackers, laptops, or mobile phones used for data collection) or voluntary termination by participants. Second, although participants performed well during the experiment, the data quality could be compromised due to undetectable sensor issues, resulting in a low tracking ratio (the percentage of eye-tracking data successfully tracked during the experiment). For instance, if the tracking ratio falls below a certain threshold, such as 75\%, the data may be considered incomplete and low-quality and can be excluded from analysis. However, the specific threshold for tracking ratio exclusion may vary depending on the experiment's context and the experimenter's decision. Figure~\ref{fig:csvmissing} shows part of an example of invalid eye-tracking data containing a lot of missing signals collected from a Tobii remote eye tracker that is formatted in CSV. Figure~\ref{fig:pupilsignal} displays an example of incomplete eye-tracking data, using pupil signals as an example. 

\textbf{Excluding invalid trials.} In eye-tracking experiments featuring multiple trials or tasks, the experimenters should check the participants' task performance closely and exclude data from invalid trials before proceeding with further data analysis. Such invalid trials include cases where participants fail to follow task instructions adequately, exhibit erratic eye movements, or engage in behaviors that interfere with the data quality. For example, participants may become distracted during a task, leading to inconsistent gaze patterns or prolonged periods of inattention. Additionally, instances of participant fatigue or discomfort can impact the quality of the eye-tracking data, as they may result in decreased focus or the adoption of unnatural viewing behaviors. Furthermore, trials affected by external factors such as environmental distractions or interruptions may also fall into this category, as they can introduce noise or bias into the collected data. Experimenters can identify invalid trials occurring under these conditions by monitoring participants' performance during the experiments or by conducting post-hoc checks, such as reviewing video recordings of the experiments or gaze recordings provided by eye-tracking analysis software (e.g., Tobii Pro Lab~\citep{tobii_pro_lab_html}. 

Building upon the initial cleaning of datasets conducted in the first step, the subsequent step in data processing mainly involves cleaning the remaining data to address any artifacts or inconsistencies within the eye-tracking data. This process consists of several aspects, including smoothing the data to eliminate high-frequency noise or fluctuations and utilizing interpolation methods to fill in missing data points or gaps in the dataset. We provide more details in Section~\ref{sub:pupil} for pupil diameter. Figure~\ref{fig:pupilsignal} demonstrates the smoothing and interpolation process applied to pupil signals as an example. 

\textbf{Smoothing.} Smoothing techniques in eye-tracking data processing aim to reduce the impact of high-frequency noise or abrupt fluctuations in the eye movement signals, thereby enhancing the clarity and interpretability of the underlying gaze patterns. One commonly used smoothing method is the moving average filter, where each data point is replaced with the average value of neighboring data points within a specified window. This averaging process helps to suppress rapid variations in the data, resulting in a smoother trajectory of eye movements over time.

\textbf{Interpolation.} Interpolation methods are invaluable in eye-tracking data processing, particularly for addressing missing data points or temporal gaps in the dataset, facilitating a more comprehensive analysis of participants' gaze behavior. For instance, when applied to pupil diameter data, interpolation techniques aim to estimate the values of missing or corrupted pupil diameter measurements by inferring them from neighboring valid data points. Commonly used interpolation methods include linear interpolation, nearest neighbor interpolation, and weighted average interpolation. When selecting interpolation methods for eye-tracking data processing, researchers should consider several factors to ensure the appropriate technique is applied based on the dataset's characteristics and the research objectives. 

\subsection{Data Segmentation}
Following data cleaning, the next step in the data processing pipeline is data segmentation, which involves dividing the continuous stream of eye-tracking data into meaningful segments or epochs based on specific criteria or events of interest. These segments serve as the basis for further analysis and interpretation of participants' gaze behavior. More details about data segmentation are provided in the following for the eye-tracking studies. 

\textbf{Segmentation criteria.} Researchers establish the criteria for segmenting the eye-tracking data based on the research objectives and experimental design. Segments can be defined by temporal factors, such as time intervals corresponding to different experimental conditions or task phases, or by event-based triggers, such as stimulus onset or participant responses. For example, in a reading task, segments could be defined based on temporal factors corresponding to different phases of the reading process, such as segmenting data according to paragraphs, sentences, or specific words of interest to the researcher.

\textbf{Segmentation method.} Different methods can be employed to segment eye-tracking data, depending on the nature of the study and the characteristics of the data. Time-based segmentation involves dividing the data into fixed or variable time intervals, while event-based segmentation relies on detecting specific events or triggers within the data stream. Hybrid approaches may combine both time-based and event-based criteria for segmentation. For example, in a video-watching task, researchers may employ different segmentation methods according to their research objectives: time-based segmentation to analyze participants' behavior during different video intervals and event-based segmentation to define segments based on detected scene changes within the video or hybrid segmentation approach.

\textbf{Segmentation validation.} It is essential to validate the effectiveness of the segmentation process to ensure that segments accurately capture the intended aspects of participants' gaze behavior. Validation may involve several aspects, such as visual inspection of segmented data. Researchers can visually inspect the segmented data to verify whether the defined segments align with the intended aspects of participants' gaze behavior. This involves reviewing the eye-tracking data alongside the corresponding stimuli or task events to ensure that segments capture relevant periods of interest. Secondly, the researcher may compare segmentations with external criteria or annotations to validate their accuracy. For example, researchers may compare segment boundaries with predefined task events or stimulus timestamps if the study involves analyzing gaze behavior during specific task phases or stimulus presentations.

\textbf{Segmentation tools.} Researchers may utilize specialized software tools or programming scripts to automate the segmentation process and facilitate efficient processing of large datasets. These tools often provide features for defining segmentation criteria, applying segmentation algorithms, and visualizing segmented data for inspection and validation (e.g., Tobii Pro Lab~\citep{tobii_pro_lab_html}). 

\subsection{Feature Extraction}
The raw eye-tracking data typically consists of frame-by-frame sensor recordings captured by the eye tracker. To gain insights into participant's visual perception, in the next step, eye-tracking metrics must be extracted from this raw data through different data processing steps. Thus, feature extraction is an essential step in the data processing pipeline, where relevant information is extracted from the cleaned and segmented raw eye-tracking data. This section discusses methods to identify eye-tracking measures, including fixations, saccades, pupil diameter, and associated statistical metrics. 

\textbf{Eye movement event detection.} It is commonly achieved through two main methods. Firstly, utilizing eye-tracking software provided by the eye tracker manufacturer, such as Tobii Pro Lab, allows for the easy extraction of eye movement events like fixations and saccades (refer to the definitions of these eye movement events in Section~\ref{sub:eventdetection}). These software packages often include built-in functions specifically designed for this purpose, simplifying the extraction process with just a few clicks. Alternatively, in cases where dedicated eye-tracking software is not available, such as in cases where an eye tracker is integrated into virtual reality (VR) headsets, eye movement events may need to be extracted by implementing custom processing pipelines. 
This custom extraction process involves carefully analyzing the raw eye-tracking data to identify and mark relevant eye movement events, such as fixations and saccades, based on predefined criteria or algorithms. The detailed algorithms for manual feature extraction can be found in Section~\ref{sub:eventdetection}.

\textbf{Statistical metrics.} Once features for fixations, saccades, and pupil diameter are extracted, researchers can compute various statistical metrics to characterize participants' gaze behavior quantitatively. These metrics offer valuable insights into various aspects of eye movement patterns and pupil dynamics observed during the specific eye-tracking task. Common statistical metrics include measures of central tendency, such as the mean and median, which provide information about the typical or average values of the data. For example, the mean fixation duration can indicate the average duration of fixations across participants or experimental conditions. Measures of dispersion, such as the standard deviation and variance, help assess the spread or variability of the data around the central tendency. For instance, the standard deviation of saccade amplitudes can indicate how much individual saccade amplitudes deviate from the mean amplitude, providing insights into the consistency or variability of saccadic eye movements during the whole task.

In conclusion, the data processing pipeline outlined in this section serves as a fundamental framework for handling raw eye-tracking data collected from participants in user studies. By systematically implementing data cleaning, filtering, and feature extraction techniques, researchers can ensure that the data is prepared in a standardized manner for further analysis. Potential artifacts and inconsistencies within the dataset are addressed through these steps, and relevant eye-tracking metrics are extracted to accurately characterize participants' gaze behavior. By following this structured approach, researchers can enhance the reliability and validity of their findings, ultimately contributing to a deeper understanding of human visual perception and cognition. 

\section{Basics Eye Movements: Fixations, Saccades, and Beyond}
This section introduces the reader to basic eye movement types, including fixations, saccades, blinks, smooth pursuits, and pupillary dynamics. In addition, we discuss how these eye movements are identified and how these data are processed. 

\subsection{Fixations}
\label{subsec_fixations}
Fixations are periods during which the eyes remain still and focus on a particular point of interest~\citep{kasneci2014applicability, tafaj2012bayesian}. Humans often acquire new information about the presented stimulus during these periods~\citep{hessels_etal_2018}. In addition, apart from knowledge intake, they are engaged with the fixated region of the stimulus. Such knowledge intake or engagement often lasts between 100 ms to 350 ms~\citep{fixsaccprotocol, rayner1998eye}, despite shorter fixation durations that have been considered in the literature. 

While fixations can be analyzed individually, one common analysis technique in the eye-tracking literature is aggregating them and analyzing their summary statistics per stimulus or task. Fixation duration and number of fixations are two important and widely used measures for assessing visual attention processes and cognitive load. For instance, longer fixation durations are typically associated with increased cognitive load or difficulty in processing information, similar to the relationship between task difficulty and fixation durations~\citep{taskdiffonvisualsearch, gao_eta_2021_chi}. However, studies have also shown that fixation behavior can vary depending on the observer's level of expertise \cite{gegenfurtner2011scan, kuebler2015exp}. In particular, it was found that although experts tend to have more fixations, these are mainly concentrated in areas with relevant information \cite{nora2017teach, gegenfurtner2011scan}. Experts also showed a shorter total fixation duration compared to novices \cite{nora2017teach}.

\subsection{Saccades}
\label{subsec_saccades}
Another common and extensively studied type of eye movement is saccadic behavior. Saccades are high-speed and ballistic eye movements that direct visual attention from one fixation to another~\citep{Agtzidis_etal_2019MM}. Like fixations, researchers often analyze aggregated measures for saccades, such as saccade velocities and amplitudes. Depending on the stimulus, higher saccade velocities may provide insights into the efficiency of the visual system in processing information and its prioritization of different regions of interest. In addition, larger saccade amplitudes may indicate that attention is being drawn from a distance~\citep{ETinwebsearch, gao_eta_2021_chi}. 

\subsection{Smooth Pursuits}
Smooth pursuit is a type of eye movement that occurs when tracking a moving object to maintain its position on the fovea~\citep{neuroscience_2nd_ed}. Unlike fixations and saccades, where the gaze is fixed or rapidly shifts between points of interest, smooth pursuit involves a slow and continuous movement of the eyes to track the moving target. This eye movement is commonly observed during activities where a moving object is present, such as driving, sports, following a conversation during mobile settings, or when encountering dynamic visual stimuli~\citep{santini2016bayesian,kasneci2015online}. Like fixations and saccades, analyzing smooth pursuit provides valuable insights into human visual perception and attention. 

\subsection{Blinks}
Blinks refer to the rapid closing and opening of the eyelids, typically lasting for a fraction of a second~\citep{schiffman2001sensation}. Apart from serving essential functions such as protecting the eyes from external factors, blinks temporarily render users blind. Thus, ignoring blinking behaviors in eye movement analysis could degrade the quality of data analysis. Blinking periods should either be excluded from the analysis or incorporated with smoothed and cleaned measurements to ensure a legitimate understanding of visual attention, perception, and cognition. On the other side, the blink rate has also been investigated in association with increased cognitive load, hence robustly detecting blinks in the eye-tracking data can reveal further insights into visual intake and cognitive processes~\citep{baccour2019camera, appel2018cross, appel2021cross, sycblinkscl14, BIONDI2023103867}. 

\subsection{Fixation and Saccade Detection: Basics}
\label{sub:eventdetection}
Fixations and saccades together form the visual scanpath. Detecting and separating these eye movements form an important step for further analysis, and there are different algorithms in the literature to achieve these, from rather lightweight, threshold-based algorithms~\citep{Agtzidis_etal_2019MM, Gao_Etal_2023} to probabilistic~\citep{tafaj2012bayesian, santini2016bayesian} or deep-learning algorithms~\citep{Zemblys2019, Elmadjian2023}. In this tutorial, we focus on two simple algorithms that are commonly used based on velocity and dispersion thresholds: Identification by Velocity-Threshold (I-VT) and Identification by Dispersion-Threshold (IDT)~\citep{fixsaccprotocol}. Both algorithms rely on the fact that fixations are the eye movements that stabilize over a region of interest, and saccades are characterized by their high speeds as defined in Sections~\ref{subsec_fixations} and~\ref{subsec_saccades}, respectively. For the algorithms that are different than I-VT and I-DT or for the customized ones such as for virtual and augmented reality (VR/AR), we refer the reader to papers by~\citep{fixsaccprotocol},~\citep{Agtzidis_etal_2019MM},~\citep{gao_eta_2021_chi}, and~\citep{bozkir2023eyetracked}. 

The I-VT algorithm considers gaze velocities while detecting fixations and saccades. After a predetermined threshold (e.g., 30-50 degrees/second), point-to-point velocities considering each sample are calculated. If the velocity of the corresponding point is below the predetermined threshold, it is labeled as a fixation point; otherwise, it is identified as a saccade point. Later, labeled fixation points are grouped together, discarding the saccade points. Then, the centroid of each fixation group is calculated, and all fixations are returned. It is also essential to ensure a minimum duration for fixations to count them as valid ones. The summary of the aforementioned process is simplified for better understanding and depicted in Algorithm~\ref{lbl_ivt}.

\begin{algorithm}
\caption{I-VT Algorithm~\citep{fixsaccprotocol}.}
\label{lbl_ivt}
\begin{algorithmic}
\Require $V_{threshold}, X_{n}$
\State $V_{t} \gets \Big((X_{t} - X_{t-1})/ \Delta t\Big)  \Big|_{t=1}^{n}$ \Comment{{\footnotesize Point-to-point velocities.}}
\While{$i \leq n$}
\If{$V_{i} < V_{threshold}$}
    \State $Type_{X_{i}} \gets fixation$
\Else
    \State $Type_{X_{i}} \gets saccade$
\EndIf
\EndWhile
\State $X_f \gets filter\_fixation\_points(X_{n}, Type_{X_{n}})$
\State $X_c \gets collapse\_conseq\_fixation\_points(X_f)$
\State $\widehat{X}_c \gets calculate\_fixation\_centroids(X_c)$
\State return $\widehat{X}_c$ \Comment{{\footnotesize Final fixation set.}}
\end{algorithmic}
\end{algorithm}

The I-VT algorithm can be considered simple, yet computationally efficient. However, one should note that different velocity and duration thresholds have been used in the literature, and suitable thresholds should be decided based on factors such as the experimental setup and the nature of the visual stimuli. 

Another commonly used algorithm for eye movement identification is the I-DT algorithm, and it is based on the idea that fixations have very small variations in gaze positions, whereas the regional movements of the saccades are large. In this algorithm, as a first step, like I-VT, a predetermined dispersion threshold (i.e., $D_{threshold}$) and a minimum duration threshold are set. Then, from a moving window starting from the first sample based on the duration threshold considering the sampling frequency of the eye tracker, the algorithm checks whether the points in the window lie within the predetermined dispersion threshold. If the calculated distance (i.e., $D = [max(x) - min(x)] + [max(y) - min(y)]$) is greater than the predetermined dispersion with $D > D_{threshold}$, it means that the points in the window do not represent a fixation together. If the calculated distance is less than the dispersion threshold, this means that points within the window are part of a fixation. In this case, the length of the window is enlarged until $D > D_{threshold}$. When the final window size is found, points within the final window are assigned to fixation by calculating the centroid of all fixation points in that window. This process is iterated until there are no more samples to evaluate. 


\begin{algorithm}
\caption{I-DT Algorithm~\citep{fixsaccprotocol}.}
\label{lbl_idt}
\begin{algorithmic}
\Require $D_{threshold}, T_{threshold}, X_{n}$
\While{$i \leq n$}
\State $W_t \gets T_{threshold} + \epsilon$
\State $X^{w}_m \gets filter\_points\_using\_window(X_n, W_t)$
\State $D \gets [max(X^{w}_m.x) - min(X^{w}_m.x)] + [max(X^{w}_m.y) - min(X^{w}_m.y)]$
\If{$D \leq D_{threshold}$}
\While{$D \leq D_{threshold}$}
    \State $W_t, X^{w}_m \gets enlarge\_window(X^{w}_m)$
\EndWhile
\State $X_c \gets calculate\_fixation\_centroid(X^{w}_m)$
\State $\widehat{X_f} \gets update\_fixation\_list(X_c)$
\EndIf
\EndWhile
\State return $X_f$ \Comment{{\footnotesize Final fixation set.}}
\end{algorithmic}
\end{algorithm}

Like the I-VT, the I-DT algorithm can also be considered simple and efficient. However, one should also adjust the dispersion and duration thresholds according to factors such as experimental design and visual stimuli. In case the reader is interested in using algorithms that adapt their parameters online (e.g., for tasks that change dynamically) and other alternatives, such as utilization Kalman filters, we refer to algorithms presented by~\citep{tafaj2012bayesian, santini2016bayesian} and~\citep{koh_Etal_2009, koh_etal_2010, Komogortsev_etal_2010}, respectively. For machine or deep-learning-based algorithms, where events are extracted in an end-to-end fashion (after adequate training of the underlying models), we refer the reader to some recent approaches, such as~\citep{fuhl2018histogram, fuhl2021fully, hoppe2016end, Zemblys2019, Elmadjian2023}. Both I-VT and I-DT algorithms are available in our eye-tracking tutorial repository\footnote{\url{https://gitlab.lrz.de/hctl/Eye-Tracking-Tutorial}}.

\subsection{Analysing Pupillary Information} 
\label{sub:pupil}
Pupil dilations and constrictions are distinct from conventional eye movements such as fixations and saccades, focusing on changes in pupil size rather than the sequence of gaze points. Pupil size is often associated with factors such as task difficulty~\citep{beatty1982task} and is related to cognitive load and mental effort~\citep{appel2021cross,castner2020pupil}. For instance, larger pupil size is typically associated with higher cognitive load~\citep{bozkir_2019_sap} and mental effort~\citep{chen_epps_2011}. However, this measure is sensitive to external influences, particularly illumination changes, especially in the wild. Therefore, careful processing is essential to interpret pupil diameter data accurately. 

To interpret the pupil sizes, once the pupil is detected and its size is identified, a processing pipeline is needed to handle the temporal aspects of the pupil data before conducting further analysis. While more complex analyses can be applied to the pupil size data, we focus on the basic preprocessing steps, including data smoothing (e.g., to eliminate blinks and measurement noise) and baseline correction, which is similar to data normalization. 

Blink removal and measurement noise reduction can be achieved through various techniques, including interpolation, averaging, and filtering. Interpolation involves filling in missing data points during blink periods with estimated values derived from surrounding data points. In contrast, averaging techniques involve replacing blink periods with the average pupil sizes before and after the blinks. Filtering involves applying a low-pass filter to the pupil data to smooth it and remove any high-frequency noise that is caused by blinks.

While such techniques can be effective in artifact removal, each of them has advantages and disadvantages. Interpolation, for instance, can introduce artifacts and distortions in the pupil data, especially if the duration of the blink is long. Filtering, on the other hand, can effectively remove blink-related noise but can also filter out some parts of the valid pupil size. Similar techniques also exist for smoothing the pupil diameter to discard outliers. This can be achieved by averaging the pupil diameter values across fixed time windows (i.e., simple moving average) or by replacing pupil diameter values with the median value within a time window (i.e., median filtering). Similar to the aforementioned eye movement identification algorithms, when choosing a method to remove blinks or smooth pupil diameter values, the experimental design and visual stimulus details should be considered to increase the quality of pupil data quality and analyses. 

In addition to the blink removal and smoothing of the pupillometry data, baseline correction is another important step, especially for normalizing the pupil size signal using a baseline duration. Baseline correction is a technique used to account for the baseline size of pupil diameter to isolate differences related to individuals. There are two mainstream techniques for baseline correction: subtractive and divisive baseline correction~\citep{Mathot2018}. In both techniques, typically, a baseline duration of up to 1 second is selected. In subtractive baseline correction, the median (or mean) baseline pupil diameter value is subtracted from each data point, resulting in positive and negative changes relative to the baseline value. In contrast, in divisive baseline correction, each pupil diameter value is divided by the median (or mean) baseline value. Using median values for baseline correction is often preferred as median values are less sensitive to noise in the data than mean values. 

\section{Visual Scanpaths: Extraction, and Comparison}
\label{sec:vis_scanpath}

Scanpaths are eye movement patterns formed from a spatiotemporal sequence of eye movements primarily consisting of fixations connected together by saccades evoked when a subject is performing certain visual tasks. Coined by \citep{noton1971scanpaths, noton1971scanpaths2}, they proposed that visual perception and the formation of viewing behavior patterns are controlled by an internal cognitive representation and that visual memory is encompassed by the visual features observed by the viewer as well as the sequence in which these features are perceived. The proposal of this theory led to decades of research on scanpath analysis and the creation of various scanpath comparison techniques. Such forms of analyses have been used to provide insight into complex cognitive processes that are otherwise not easily extracted from the subject. This has been used in a wide variety of applications such as usability design \citep{ichindelean2021comparative}, expertise classification \citep{castner2020deep}, neurological disorders \citep{li2020eye}, and educational research \citep{popelka2022scanpath}. 

While other eye movements such as smooth pursuit, microsaccades, and ocular drifts exist, they are often overlooked due to challenges in isolating them from the eye-tracking signal. Scanpaths typically prioritize fixations and saccades, although some scanpath comparison methods tend to discount saccades due to the suppression of visual perception during this event \citep{kubler2017subsmatch}. There are different approaches to comparing scanpaths. In general, a scanpath comparison metric produces a small distance between scanpaths that are similar in terms of shape and timing and a large distance otherwise. Common ways to represent them are in the form of strings, probabilistic models, or geometrical vectors--each having different comparison methods. 

\subsection{String Alignment Methods}
In this approach, a scanpath is typically represented by encoding the spatial location information of fixations into a string, where each character denotes the region of interest (ROI) the gaze fixates upon. Various string-comparison algorithms can be applied to these encoded strings, such as measuring pairwise string similarity through methods like string edit distance. One such algorithm is the Hamming Distance, which counts the differing characters at corresponding positions in two strings of equal length. However, the Hamming Distance is limited due to its restrictive nature, as it only allows substitutions and requires identical string lengths. An alternative algorithm, ScanMatch \citep{cristino2010scanmatch}, offers greater flexibility albeit with increased complexity.

\subsubsection{ScanMatch} 

\begin{figure}[b]
\centering
\includegraphics[width=0.6\linewidth]{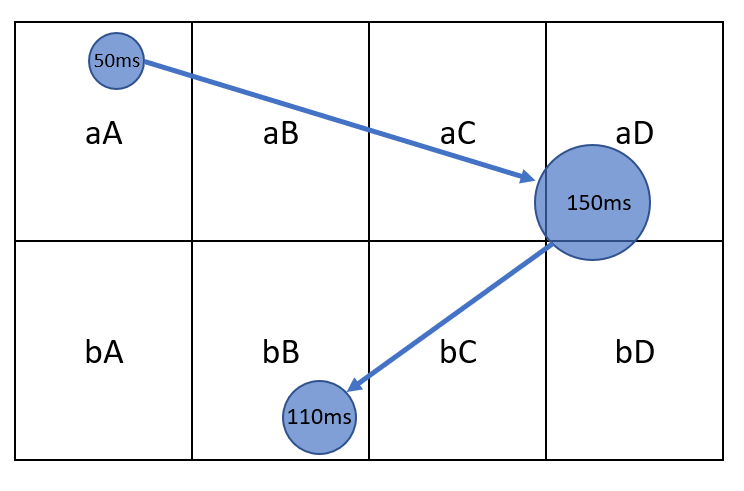}
\caption{A stimulus divided into $4 \times 2$ bins. The fixations are represented by circles of varying radius, representing their corresponding durations in milliseconds. The arrows between subsequent fixations represent the saccades.}
\label{fig:scanmatch}
\end{figure}

This method \citep{cristino2010scanmatch} utilizes an established algorithm used for comparing DNA sequences in bioinformatics, known as the Needleman-Wunsch algorithm \citep{needleman1970general}. In the context of scanpath comparison, the scanpath is spatially and temporally binned to encode a string that preserves fixation location, duration, and order information. This allows for comparing spatial, temporal, and sequential characteristics between scanpaths. Before encoding, the visual stimulus (e.g., the image being viewed by the subject) is divided into ROIs. Each ROI is assigned a letter (or a group of letters if there are more than 26 ROIs in the image). In Figure \ref{fig:scanmatch}, a normal string sequence of ``aAaDbB'' can be derived from the given scanpath without temporal binning. With temporal binning of 50 ms bins, the resulting encoding is ``aAaDaDaDbBbB.''  

The scanpath encodings can then be pairwise compared by finding the optimal alignment where, for each position, the following cases may occur: 
\begin{enumerate}
    \item The pair is a match. 
    \item A gap is inserted in sequence 1. 
    \item A gap is inserted in sequence 2.
    \item The pair is a mismatch (substitution).  
\end{enumerate}
This optimal alignment can be calculated efficiently using the Needleman-Wunsch algorithm, a dynamic programming approach~\cite{lew2006dynamic}, as described in Algorithm \ref{algo:nmw}.
Only ``two'' parameters need to be set: the substitution matrix and the gap penalty.
This algorithm introduces the following concepts: 

\begin{enumerate}
    \item \textbf{Substitution Matrix}: Provides the score for substituting one letter with another. The higher the score, the more similar the pair of letters are to each other. The scores reflect the relationships between ROIs and can be based on distance (how near they are to each other), color, and semantic segmentation. Moreover, a cutoff point needs to be chosen, determining whether the score between two ROIs should be positive (highly related) or not (loosely related). 
    \item \textbf{Gap Penalty}: This is incurred when an element in the sequence is aligned with a gap instead of a substitution. Depending on how the penalty is determined, the gap penalty can encourage (or discourage) gaps over substitutions. 
    \item \textbf{Scoring}: As the alignment score is highly affected by sequence lengths, the score needs to be normalized, with the highest score being 1, as shown in Eq. \ref{eq:NWnorm}.
    
    \begin{figure*}[!ht]
        \begin{equation}
        \text{Normalized score} = \frac{\text{score}}{\text{Max (substitution matrix)} * \text{length of longest sequence} }
    \label{eq:NWnorm}
    \end{equation}
    \end{figure*}
    
\end{enumerate}

\begin{algorithm}
\caption{Needleman-Wunsch Algorithm with Insertion and Deletion Costs~\citep{needleman1970general}.}\label{needleman-wunsch-insertion-deletion}
\begin{algorithmic}[1]
\Procedure{NeedlemanWunsch}{$seq_1, seq_2, S, c_{ins}, c_{del}$}
    \State $m \gets$ length of $seq_1$
    \State $n \gets$ length of $seq_2$
    \State Create a 2D matrix $DP$ of size $(m+1) \times (n+1)$
    
    \For{$i \gets 0$ to $n$}
        \State $DP[i][0] \gets i \cdot c_{del}$ \Comment{Cost of deletion in $seq_1$}
    \EndFor
    
    \For{$j \gets 0$ to $m$}
        \State $DP[0][j] \gets j \cdot c_{ins}$ \Comment{Cost of insertion in $seq_2$}
    \EndFor
    
    \For{$i \gets 1$ to $m$}
        \For{$j \gets 1$ to $n$}
            \State $matchScore \gets DP[i-1][j-1] + S(seq_1[i], seq_2[j])$ \Comment{Match/Mismatch cost}
            \State $deletionScore \gets DP[i-1][j] + c_{del}$ \Comment{Cost of deletion in $seq_1$}
            \State $insertionScore \gets DP[i][j-1] + c_{ins}$ \Comment{Cost of insertion in $seq_2$}
            \State $DP[i][j] \gets \min(matchScore, insertionScore, deletionScore)$ \Comment{Fill DP matrix}
        \EndFor
    \EndFor
    
    \State \Return $DP[n][m]$ \Comment{Final alignment cost}
\EndProcedure
\end{algorithmic}
\label{algo:nmw}
\end{algorithm}

String-based methods have their limitations, particularly with the need to predefine ROIs, which can constrain the stimulus and potentially compromise basic spatial information about the scanpath \citep{jarodzka2010vector}. Furthermore, fixations occurring near the borders of these predefined ROIs may not be accurately quantified. Additionally, such methods often fail to preserve the inherent locality of the data \cite{newport2022softmatch}. 

\subsection{Geometrical Methods} 
Another approach to comparing scanpaths involves their geometrical properties. This method defines the problem as finding the optimal mapping between fixation locations in both scanpaths according to their spatial distance with neighboring fixations. \citep{mannan1995automatic} proposed a nearest neighbor approach in which the eye movement is represented as a set of fixations in the form of x and y coordinate pairs. Each fixation in one set is mapped to the nearest fixation from the other set, resulting in a set of mapping distances. The sum of all mapping distances is then calculated after normalization, accounting for the length of eye movement sequences.

\begin{figure}[b]
\centering
\includegraphics[width=0.5\linewidth]{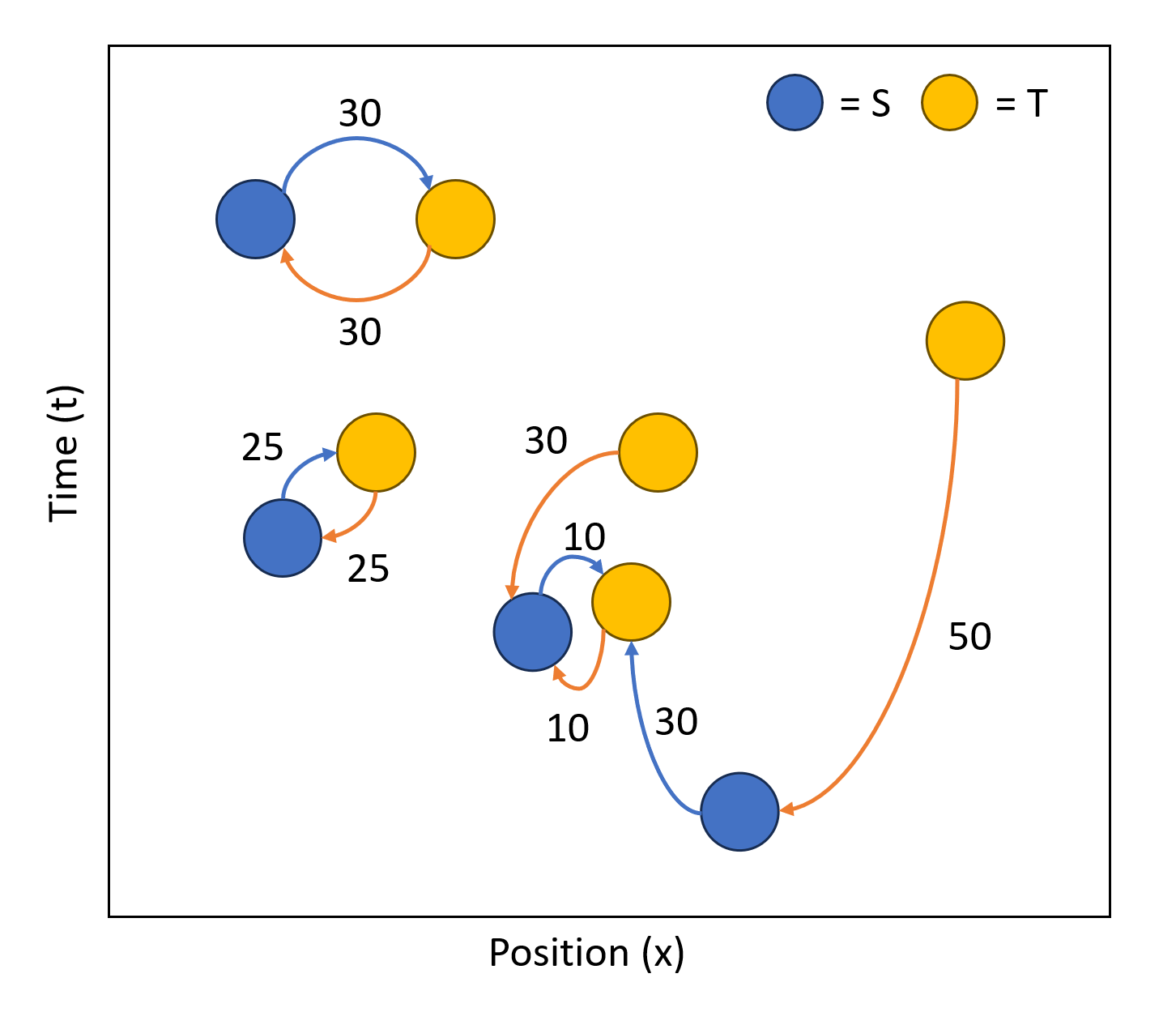}
\caption{A simplistic example of the Eyenalysis approach proposed by \citep{mathot2012simple}. The resulting distance is calculated as follows: $D(S, T) = ((30 + 25 + 10 + 30) + (30 + 25 + 30 + 10 + 50))/max(4, 5) = 48$.}
\label{fig:eyenalysis}
\end{figure}

\citep{mathot2012simple} extends this approach by proposing a method called Eyenalysis that makes use of a double mapping technique, which involves mapping each fixation (represented as a vector of arbitrary dimension containing various properties such as spatial and temporal information) in one scanpath to its nearest neighbor in the other, and then repeating the process the other way around. This foregoes the decision step of whether fixations should have multiple connections with the other scanpath. Figure \ref{fig:eyenalysis} exemplifies the double mapping technique. The distance is calculated by summing up the point-mappings and normalizing by the length of the longest scanpath. The mapping between points $p$ from scanpath $S$ and $q$ from scanpath $T$ is associated with the Euclidean distance $d(p, q)$ as defined below (where $n$ is the number of dimensions): 
\begin{equation}
        d(p, q) = \sqrt{\sum_{i=1}^n{(p_i - q_i)^2}}
\label{eq:eyen_pointdist}
\end{equation}

The sequence mapping $D(S,T)$ between $S$ and $T$ is then calculated as follows: 
\begin{equation}
        D(S, T) = \frac{    \sum^{n_S}_{i=1}{d^i_S} + \sum^{n_T}_{j=1}{d^j_T}  } {max(n_S, n_T)}
\label{eq:eyen_pointdist1}
\end{equation}
    
where $n_S$ and $n_T$ are the lengths of $S$ and $T$ respectively. The numerator refers to the sum of all point mappings from $S$ to $T$ and from $T$ to $S$, respectively. 

Another approach is MultiMatch \citep{dewhurst2012depends}, which is a more complex vector-based method that produces scanpath distances across multiple dimensions while still preserving positional and temporal information. First, a process of amplitude-based and direction-based simplifications on the scanpath is performed by grouping small, locally contained saccades together through thresholding and merging successive saccades following the same general direction. Representative values such as location and fixation duration are chosen as vector dimensions, and an optimal mapping is determined using the Dijkstra algorithm. 

\subsection{Probabilistic Methods} 
\begin{figure}[t]
\centering
\includegraphics[width=0.8\linewidth]{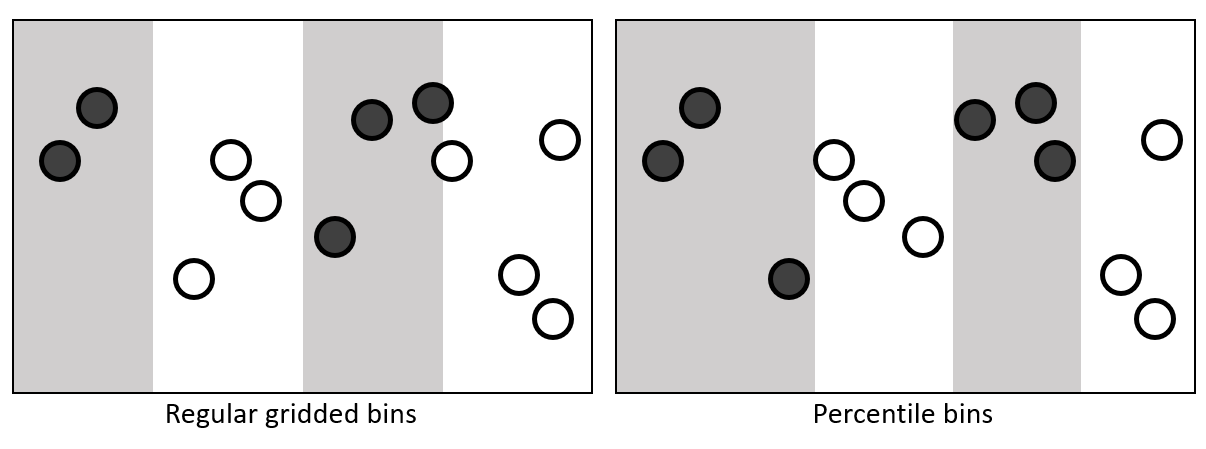}
\caption{Two ways of determining ROIs as presented by \citep{kubler2017subsmatch}. On the left, regular gridded bins are formed and overlayed over the data. On the right, the bins are formed using data percentiles. The fixations are assigned to their corresponding bins represented by the dark and light regions on both figures.}
\label{fig:subsmatch}
\end{figure}

Probabilistic methods tackle the high normal variability between scanpaths by hypothesizing the stochastic nature of eye movement parameters as random variables sampled from underlying stochastic processes \citep{coutrot2018scanpath}. Hidden Markov Models (HMMs) have been used to model eye movement by learning an HMM from one or a group of scanpaths, which may incorporate dynamic and individualistic properties of the gaze behavior allowing for the extraction of patterns characteristic of a certain class \citep{coutrot2018scanpath}. Subsmatch 2.0 \citep{kubler2017subsmatch} is another probabilistic method that employs a novel string kernel approach for scanpath comparison, where the scanpaths are first encoded into a string and further spliced into smaller subsequences. The frequency of specific subsequences that resemble typical, repeatedly occurring behavioral patterns is then used as a similarity feature. The concept is derived from the transition matrix approach, which counts the number of transitions from one ROI to another. It considers exploratory gaze patterns that consist of sequences of more than two subsequent fixations, using n-gram features to represent subsequences of length n. The method can be applied to scenarios where labeling ROIs is not possible, such as viewing abstract art or interactive, dynamic scenarios, using a regular grid or percentile mapping to determine ROIs from the data as shown in Figure \ref{fig:subsmatch}. It can also infer scanning patterns associated with specific experimental factors by applying machine learning techniques, such as a support vector machine (SVM) with a linear kernel. 

\subsection{Deep Learning Methods} 

\begin{table}[t]
\centering
\caption{Comparison of Different Approaches to Scanpath Analysis}
\label{table:scanpath-comparison}
\begin{tabular}{>{\centering\arraybackslash}p{0.15\linewidth}p{0.25\linewidth}p{0.25\linewidth}p{0.25\linewidth}}
\toprule
\multicolumn{1}{c}{\begin{tabular}[c]{@{}c@{}}\textbf{Scanpath}\\ \textbf{Comparison}\\ \textbf{Method}\end{tabular}} & \multicolumn{1}{c}{\textbf{Characteristics}} & \multicolumn{1}{c}{\textbf{Advantages}} & \multicolumn{1}{c}{\textbf{Limitations}} \\
\midrule
\textbf{String Alignment} & 
Encodes spatial locations of fixations into strings for comparison using algorithms like the Hamming Distance and ScanMatch. &
Applicable for comparing spatial, temporal, and sequential characteristics of scanpaths. & 
Requirement of predefined ROIs may compromise spatial information. Inherent locality is also not preserved. \\
\midrule
\textbf{Geometrical} & 
Captures the optimal mapping between fixation locations based on spatial distance, employing techniques like nearest neighbor mapping and Eyenalysis for double mapping. &
Does not require predefined ROIs and more properly preserves spatial information compared to string alignment methods. &
May not effectively capture the temporal sequence of fixations. \\
\midrule
\textbf{Probabilistic} & Models eye movements as random variables sampled from underlying stochastic processes using methods like HMMs and Subsmatch 2.0. & Accounts for the stochastic nature of eye movement, capturing dynamic and individualistic gaze behavior patterns. & Complex to implement and interpret, requiring sophisticated statistical knowledge. \\
\midrule
\textbf{Deep Learning} & Employs neural networks to learn hierarchical representations from scanpaths, capturing both spatio-temporal characteristics and semantic content. & Can automatically extract complex patterns and relationships without the need for predefined ROIs. & Requires significant computational resources and large datasets for training. \\
\bottomrule
\end{tabular}
\end{table}

\begin{figure}[b]
\centering
\includegraphics[width=0.8\linewidth]{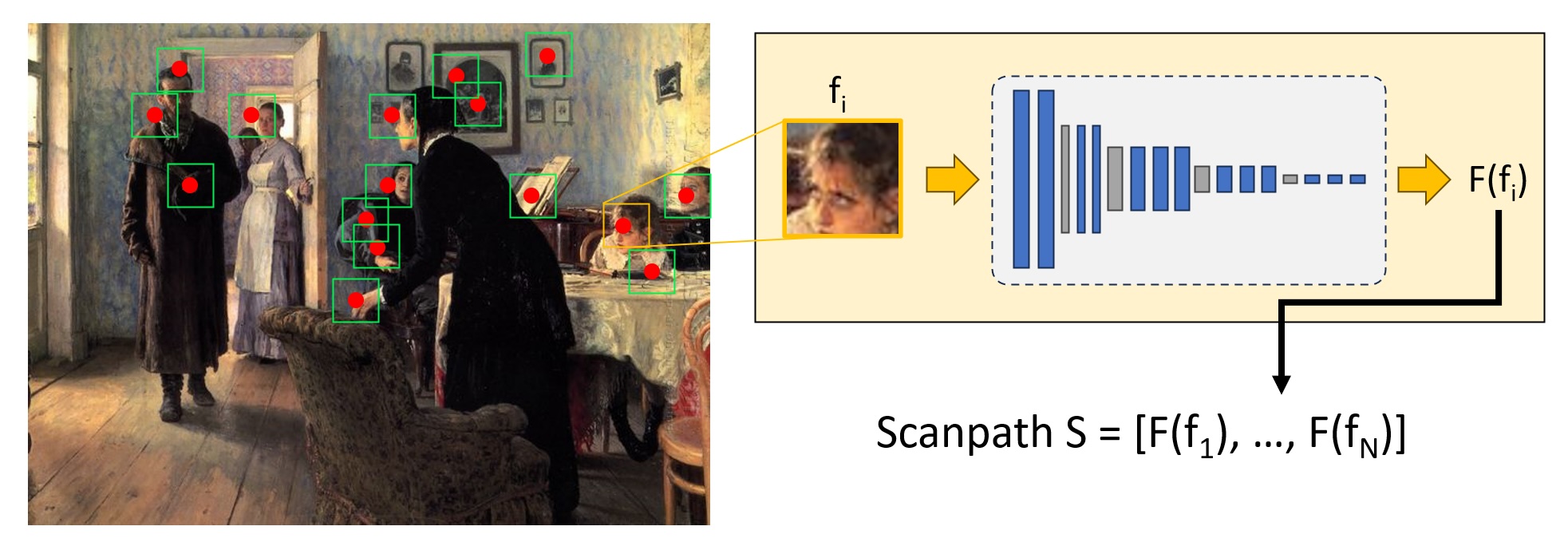}
\caption{The proposed approach by \citep{castner2020deep}. A corresponding image patch (green box) is extracted for each fixation (red dot). Each image patch ($f_i$ where $i = 1...N$ and N is the total number of fixations) is passed through a CNN to extract the features $F(f_i)$, which are concatenated into a scanpath vector S. This can then be compared with other scanpath vectors.}
\label{fig:deepscan}
\end{figure}

Deep learning algorithms have emerged as powerful tools in scanpath analysis. These algorithms, based on artificial neural networks, can automatically learn hierarchical representations of data, making them well-suited for tasks involving complex patterns and relationships. In the context of scanpath analysis, deep learning algorithms can be used to extract features from scanpaths that capture both low-level and high-level information, such as the spatial and temporal characteristics of eye movements and the semantic content of the visual stimuli being viewed. With the growing popularity of machine learning and the emergence of easily accessible models, along with the introduction of transfer learning and fine-tuning, a handful of deep learning-based scanpath comparison methods have been introduced. For instance, \citep{castner2020deep} proposed a method of extracting scene information at the fixation level by incorporating convolutional neural networks (CNN) as a means to extract features from each fixation according to the corresponding image patch on which the fixation has landed. The proposed approach is exemplified in Figure \ref{fig:deepscan}. Table~\ref{table:scanpath-comparison} shows the comparison of different approaches to scanpath analysis.

\section{Visualization of Eye-Tracking Data: Best Practices}
The benefit of visualizing eye-tracking data lies in the ability of visualization techniques to facilitate the comprehension of spatio-temporal characteristics and intricate relationships embedded within the data, thereby enhancing the interpretability of the results. Therefore, visualizing eye-tracking data is important for an effective communication of the results and the interpretation of complex patterns in eye movement behavior. This section introduces important techniques and best practices for scientifically visualizing eye-tracking data, including saliency maps, scanpaths, and gaze plots.

\begin{figure}[t]
  \centering
   \subfigure[CUB-GHA dataset.]{{\includegraphics[height=0.3\linewidth,keepaspectratio]{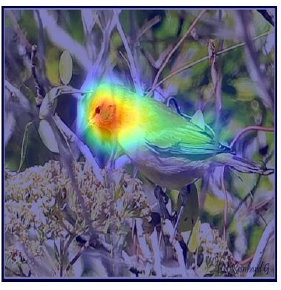}}}%
   \quad
   \subfigure[CXR-Eye dataset.]{{\includegraphics[height=0.31\linewidth,keepaspectratio]{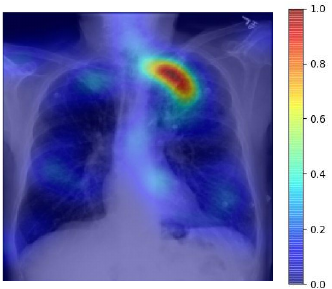} }}%
  \caption{Saliency maps on (a) CUB-GHA and (b) CXR-Eye datasets. Color red indicates more attention from humans, while color blue indicates less attention.}
   \label{fig:viz-heatmap}
\end{figure}

\subsection{Saliency Maps}

Saliency maps are among the most commonly used visualizations of eye-tracking data and serve as a powerful means to quickly and easily highlight areas within a visual scene that attract the most visual attention. These maps aggregate fixation data into a heat-like representation that intuitively illustrates the distribution of visual attention. More specifically, areas that receive a high amount of or longer fixations are typically shown as hot or red, indicating these spots are of high interest or saliency to the observers. In contrast, areas with fewer or shorter fixations appear cooler, marked by blue, suggesting that less attention was paid to these parts of the visual scene.

The typical construction of saliency maps involves plotting of fixation points over the stimulus and applying a Gaussian blur to each point, thereby creating a continuous probability distribution that represents visual attention. Due to its simplicity, saliency maps are used to quickly assess visual attention distribution in various applications, such as usability testing, marketing research, psychology and many more~\citep{borji2012state,eder2021support,kou2023advertising,yan2021review,rong2022user}.

\begin{figure}[h]
    \centering
    \includegraphics[width=.3\linewidth]{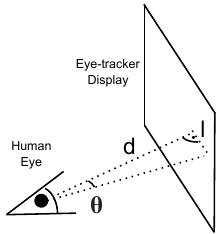} 
    \caption{Illustration of a human observing an image on the eye tracker display.}
    \label{fig:gaussian}
\end{figure}

Their construction is described in detail in the following. To visually illustrate human attention, i.e., gaze fixation, it is common to add a Gaussian filter on fixation points to form a heatmap~\citep{judd2012benchmark}, which is also called \textit{saliency} map~\citep{kummerer2016deepgaze}, which is depicted in Figure~\ref{fig:viz-heatmap}. This technique is usually applied to the gaze data remote eye trackers collect. In this section, we introduce how to visualize the saliency map for fixation in practice. We use the gaze data from the dataset CUB-GHA~\citep{rong2021human} as an example to clarify the procedure~\footnote{Code for visualization can be found at~\url{https://gitlab.lrz.de/hctl/Eye-Tracking-Tutorial}.}. Before collecting gaze data, the practitioner should gather the following information: 
\begin{itemize}
     \item Display resolution of the remote eye tracker
     \item Distance between the human (face/eyes) and the display of the remote eye tracker. 
\end{itemize}

After collecting the fixation gaze data, the saliency maps can be visualized based on the fixation position (on the image) and duration. Concretely, Figure~\ref{fig:gaussian} illustrates a human observing an image on the eye tracker display. We post-process every fixation location as a Gaussian distribution ${N}(\mu,\,\sigma^{2})$ on the gaze fixation saliency map, where $\sigma$ is 75 pixels (in the display's resolution). We calculate the standard deviation $\sigma$ as follows. 
$d$ refers to the distance between the human eye and the eye tracker display. In practice, for instance in~\citep{rong2021human}, $d$ is set to 60 $cm$, and the visual angle $\theta$ is set to 2$^{\circ}$ following \citep{vickers2007perception}. In this case, $l=\tan{2^{\circ}} \cdot d = 21$ $mm$. According to the settings of the display, in the horizontal direction, the length of the display is 530 $mm$, and the resolution is 1920 pixels. Therefore, we can get that $l=$ 21 $mm$ covers approximately 75 pixels on the display. We set 75 pixels as the standard deviation with the image rescaled to the display resolution (1920 $\times$ 1080). The saliency map is rescaled to its original size afterwards.

\subsection{Scanpaths}

\begin{figure}[t]
\centering
\includegraphics[width=0.96\linewidth]{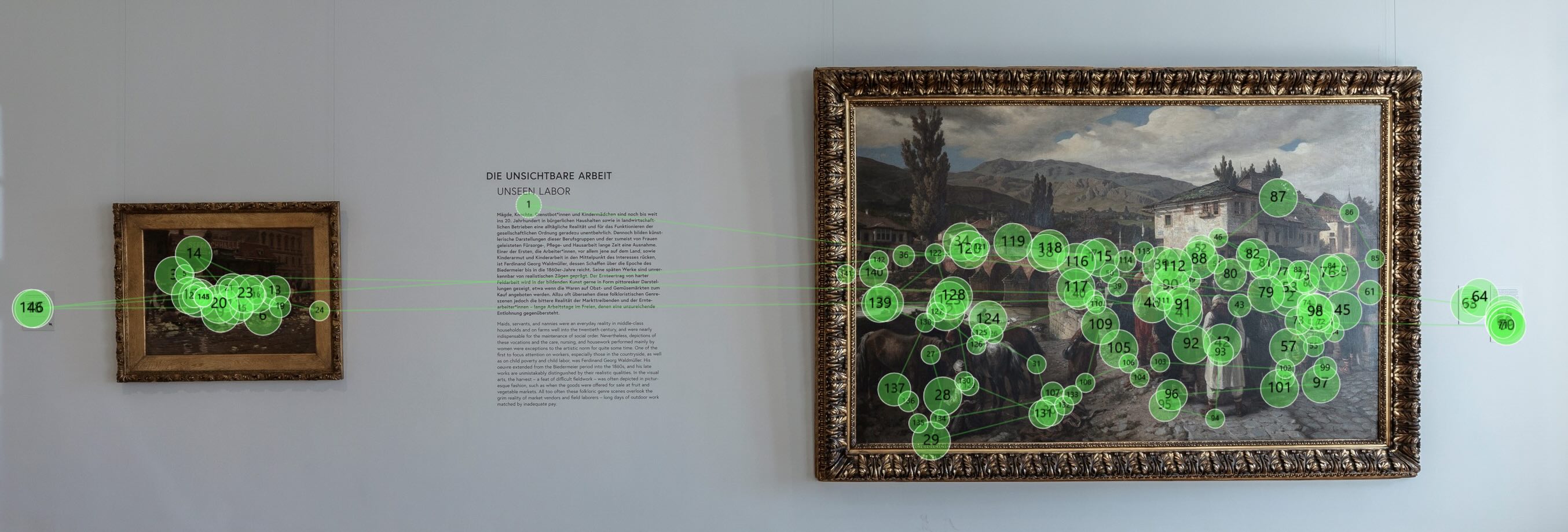}
\caption{Scanpath visualization representing measured and transformed fixations of a museum visitor viewing paintings and descriptive texts sequentially.}
\label{fig:scanpath_museum}
\end{figure}

The concept of visual scanpaths, as previously discussed, applies to the sequential order of fixations and accompanying saccadic movements \citep{goldberg2014scan, noton1971scanpaths2}. Figure \ref{fig:scanpath_museum} visually represents such a scanpath, where fixations are depicted as dots and saccades as connecting lines. The size of the dots correlates with fixation duration, providing precision to the scanpath and enabling insights into overall gaze behavior \citep{holmqvist2011eye}.

Accurately visualizing a scanpath requires essential information such as stimuli dimensions, fixation coordinates, and corresponding durations. This data extraction process is relatively straightforward with remote eye-tracking systems, involving knowledge of stimulus placement on the display and subsequent data processing for fixation metrics. However, in scenarios involving navigation through three-dimensional environments while wearing a head-mounted eye tracker, such as exploring a museum, visualizing resulting scanpaths introduces added complexity. The spatial nature of the setting necessitates careful transformation of measured fixations into a 2D representation.

A simplified approach to scanpath visualization may entail focusing solely on illustrating saccadic lines. This streamlined method proves advantageous when comparing multiple scanpaths by overlaying them on the same stimuli. By omitting dots representing fixation duration, the level of visual disorder and overload in the visualization is reduced, facilitating clearer comparisons between different scanpaths. This approach enhances the interpretability of visual exploration patterns in complex environments, thereby aiding researchers in understanding gaze behavior dynamics more effectively.

\subsection{Gaze Plots}

Gaze plot visualization is a method used to depict eye-tracking data, offering valuable insights into an individual's attentional focus and the sequence in which they explore different areas within a stimulus~\citep{burch_etal_2014, TAKAHASHI2018449, shrestha2007eye}. Unlike scanpaths, which offer a chronological perspective of eye movements, gaze plots focus on visualizing an individual's gaze path by illustrating the sequence and spatial distribution of fixations and saccades over a specified time window~\citep{TAKAHASHI2018449}. By graphically representing an individual's gaze path, gaze plots depict how attention shifts across different parts of the stimulus, aiding researchers and practitioners in understanding visual exploration patterns.

\begin{figure}[t]
\small
\centering
\includegraphics[width=.6\linewidth]{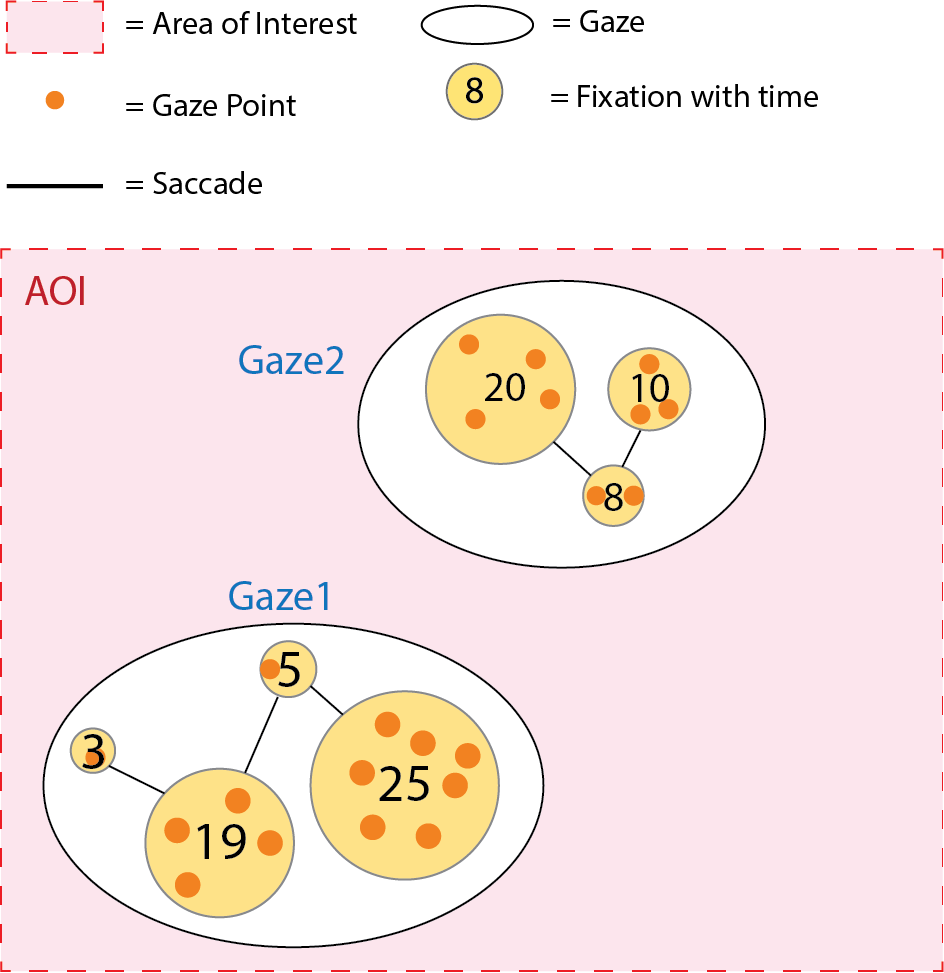}
\caption{Gaze plot visualization adapted from~\citep{Blascheck}.}
\label{fig:gaze_plot_explanation}
\end{figure}

In a traditional gaze plot, as depicted in Figure~\ref{fig:gaze_plot_explanation}, each fixation is typically represented by a circle, with the size of the circle indicating the duration of the fixation. Similarly to scanpaths, lines connecting the circles depict the saccades, or rapid eye movements, between the fixations. 

Gaze plot visualization provides valuable insights into visual attention patterns and cognitive processes. This information can be leveraged to inform the design of more effective interfaces, advertisements, or educational materials, enhancing user engagement and comprehension.

\section{Ethical Considerations}  
As eye-tracking data is collected from human participants, one should consider important ethical issues when the purpose of the data collection is research. The eye-tracking community has recently also drawn attention to such issues~\citep{10445548}, and we here cover the most relevant aspects. In general, in eye-tracking experiments, it is important to obtain ethical approval from the institutional review board (IRB) before conducting user studies to ensure the well-being of the participants. IRBs assess the ethical aspects of research and relevant experiments to ensure studies adhere to ethical standards and regulations. These ethical aspects include but are not limited to recruitment methods, informed consent procedures, risks and benefits for the participants, participants' privacy, and confidentiality of personal information. The Belmont Report~\citep{belmont_report} and the Declaration of Helsinki~\citep{Declaration_of_Helsinki_7th_ed} provide several principles and guidelines for research ethics to this end. Both of these reports also acknowledge informed consent and risk and benefit assessment as essential applications, further recognizing the importance of participant selection, privacy, and confidentiality. 

\textbf{Informed consent.} Informed consent is a critical component of any research study that includes human participants, and it is the process of obtaining voluntary agreement from a participant to participate in a study after they are provided with relevant information about the study, such as the purpose of the study, experimental procedures, potential risks, and benefits. It is essential for the researchers that the informed consent is designed in a way that participants clearly understand what they consent to, and it protects their rights and welfare. 

In a step-by-step manner, in most lab studies, the researcher should provide a description of the details of the study, equipment, procedures involved, and potential risks and benefits, in addition to the written consent form. The researchers should also provide information on how the data will be collected, stored, and used further, with the researcher's or principal investigator's contact information. The researcher should answer any questions that participants may have and ensure they understand the information. Once participants have reviewed and understood the details and consent form, they should sign it to indicate their voluntary agreement to participate in the study. If a user study involves minors who are under 18 years of age, as they can only assent but not provide consent, a parent or a legal guardian should sign the consent form. If a user study includes data collection in the form of images or videos (e.g., video data of eye images that include iris textures, which form personally identifiable data), the consent forms should explicitly mention this and provide further details on how the data will be processed, stored, and managed. In addition, the researcher should ensure that the participant is not pressured into providing consent, and they should clearly state that participants are free to withdraw from the study without any consequences at any time. The researcher should keep copies of the signed consent forms and other relevant experiment documents. 

\textbf{Risks and benefits.} Researchers and IRBs should carefully assess the potential risks and benefits that experiments provide in experiments. Most eye-tracking experiments have small to negligible risks unless the presented stimuli in the experiments are not sensitive and risky or evoke emotions that might harm the participants psychologically or physically. The experiments in extended reality (XR) might cause cybersickness, nausea, or dizziness. In addition, long eye-tracking experiments might cause boredom. 

Some studies might involve deception, making the experiments and ethical processes more complex. The user studies and experiments that include deception must be carefully designed and thoroughly reviewed by IRBs as these involve greater risks to participants than conventional experiments. While the informed consent procedures remain similar for the studies with deception, in such experiments, the participants must be informed and agree with the deception protocol in advance. This can be achieved by recruiting the participants from a pool that participants have already approved in advance to participate in experiments that involve deception. In such experiments, it is essential to fully debrief participants after the completion of the experiment. Debriefing should include an explanation and purpose of the deception. As deception can potentially cause harm to participants, it is important to consider the risks involved in the study and ensure that they are justified. Deception should only be used when it is necessary to answer a research question that cannot be answered without deception. In addition, to ensure the wellness of the participants, it may be necessary to include control conditions in the experiments to compare the outcome of the deception condition with the control condition. 

Regarding benefits, gift cards, monetary support, or hourly course credits (in some institutions) are usual ways of compensating participants' time and effort. When compensated with monetary support, researchers should keep the minimum hourly rates in the particular country in mind; however, compensations should not be very high to attract potential participants that participants only join the study due to the high amount of money, which might create a potentially biased sample pool. 

\textbf{Privacy and Confidentiality.} Regardless of the informed consent or deception processes, it is vital to maintain the confidentiality of personal information and protect participants' privacy. Collected data, including eye tracking, should be anonymized if possible, and personally identifiable information should not be included in any publications unless the participants provide explicit permission. In addition, data from user studies, such as eye movement data, can be representative of personal identifiers or sensitive participant characteristics depending on the displayed stimulus~\citep{bozkir2023eyetracked, liebling_preibusch_2014}. Researchers and practitioners must not attempt to identify such identifiers and characteristics unless the research questions require these to be carried out and participants agree with them. In addition, especially for practitioners, it is advisable that privacy-preserving approaches for eye-tracking data are employed~\citep{PPGE_etra_2020, diff_privacy_2021, 10049660, elfares2023federated, elfares2024privateyes, ozdel_etal_2024privacy_ETRA}. 

\section{Summary and Conclusion}

This tutorial represents a gentle and comprehensive introduction to the essentials of eye-tracking user studies. We began with an overview of the technology and the calibration necessary for accurate data collection. We progressed by defining the basic types of eye movements -- fixations, saccades, blinks, and smooth pursuits -- and their significance in understanding human cognitive processes. Popular algorithms for detecting these movements and the formation of visual scanpaths were presented, along with techniques for processing pupillometry data. We extended the discussion to the complexities of visual scanpaths, including their processing, comparison, and visualization techniques. The technical content was complemented and rounded up by addressing ethical considerations in eye-tracking user studies, emphasizing the importance of informed consent, the assessment of risks and benefits, as well as the safeguarding of participant privacy. 

We believe that this holistic approach to eye-tracking user studies not only enhances the students' and practitioners' understanding of eye movements in many applications and scientific fields, especially in the context of human-computer interaction, but also reinforces best practices in conducting and evaluating eye-tracking research in a principled and ethical manner.

\bibliographystyle{unsrtnat}
\bibliography{references}  






\end{document}